\documentclass[printer]{aa}

\usepackage{amssymb}
\usepackage{graphicx}
\usepackage{mathptmx}
\usepackage{amsmath}
\usepackage{color} 

\usepackage{txfonts}
\usepackage{natbib}

\def\eff{\epsilon_{ff}}
\def\tff{\tau_{ff}}
\def\rhost{\rho_{\star}}
\def\Sigst{\Sigma_{\star}}
\def\st{_{\star}}
\def\Ms{M_{\odot}}
\def\sfe{star formation efficiency }

\begin{document}

\title{Local-Density Driven Clustered Star Formation}
\author{G. Parmentier \inst{1,2,3} and 
        S. Pfalzner \inst{2} }

\institute{\inst{1} Astronomisches Rechen-Institut, Zentrum f\"ur Astronomie, Heidelberg Universit\"at, M\"onchhofstr. 12-14, D-69120 Heidelberg, Germany \\ 
           \inst{2} Max-Planck-Institut f\"ur Radioastronomie, Auf dem H\"ugel 69, 53121 Bonn, Germany \\ 
           \inst{3} Argelander-Institut f\"ur Astronomie, Bonn Universit\"at, Auf dem H\"ugel 71, D-53121 Bonn, Germany \\
           \email{gparm@ari.uni-heidelberg.de}
}
\date{Received 22 May 2012 / Accepted 23 October 2012}
%$^{2}$ Argelander-Institut f\"ur Astronomie, Bonn Universit\"at, Auf dem H\"ugel 71, D-53121 Bonn, Germany

\abstract
{A positive power-law trend between the local surface densities of molecular gas, $\Sigma_{gas}$, and young stellar objects, $\Sigma_{\star}$, in molecular clouds of the Solar Neighbourhood has recently been identified by Gutermuth et al.  How it relates to the properties of embedded clusters, in particular to the recently established radius-density relation, has so far not been investigated.}
{We model the development of the stellar component of molecular clumps as a function of time and initial local volume density.  Our study provides a coherent framework able to explain both the molecular-cloud and embedded-cluster relations quoted above.}
{We associate the observed volume density gradient of molecular clumps to a density-dependent free-fall time.  The molecular clump star formation history is obtained by applying a constant \sfe per free-fall time, $\eff$.}
{For volume density profiles typical of observed molecular clumps (i.e. power-law slope $\simeq -1.7$), our model gives a star-gas surface-density relation of the form $\Sigma_{\star} \propto \Sigma_{gas}^2$.  This is in very good agreement with the relation observed by Gutermuth et al.  Taking the case of a molecular clump of mass $M_0 \simeq 10^4\,\Ms$ and radius $R \simeq 6\,pc$ experiencing star formation during 2\,Myr, we derive what \sfe per free-fall time matches best the normalizations of the observed and predicted ($\Sigma_{\star}$, $\Sigma_{gas}$) relations.  We find $\eff \simeq 0.1$.  
We show that the observed growth of embedded clusters, embodied by their radius-density relation, corresponds to a surface density threshold being applied  to developing star-forming regions.  The consequences of our model in terms of cluster survivability after residual star-forming gas expulsion are that due to the locally high \sfe in the inner part of star-forming regions, global \sfe as low as 10\% can lead to the formation of bound gas-free star clusters.}
{}
  

\maketitle
% --------------------------------
\section{Introduction}
\label{sec:intro}
% --------------------------------
Properties of star-cluster forming regions are crucial in determining how the nascent cluster dynamically responds when the gas left unprocessed by star formation is driven out due to stellar feedback.  Does the cluster survive as a bound entity -- albeit depleted of a fraction of its stars, or is it fully dispersed into the field?  Pivotal properties influencing cluster-survival likelihood include the star formation efficiency of cluster-forming regions and their potential well depth.  The latter can be quantified through their mass-radius relation, which several authors report to scale  as $m_{ecl} \propto r_{ecl}^2$, where $m_{ecl}$ and $r_{ecl}$ are the embedded-cluster stellar mass and radius, respectively \citep{lad03,ada06}.  We stress that this relation refers to the stars only, i.e. $m_{ecl}$ does not account for the unprocessed gas.

This mass-radius relation is one of constant mean surface density and can stem from the method applied to measure the embedded-cluster radius and the stellar mass it contains.  As pointed out by \citet{all07}, if the mass and radius of a cluster are defined based on a surface density threshold, they are necessarily sensitive to the adopted density cut-off.  This also determines what mass fraction of the stars is hosted by the `halo' surrounding the star cluster, as opposed to the cluster itself.  For a given star-forming region, the higher the surface density threshold, the smaller the cluster mass and radius, the larger the surrounding `halo'.  This renders an accurate definition of the cluster-forming region problematic.  

Besides, it is worth stressing that the dynamical response of a cluster to the residual star-forming gas expulsion depends on the cluster-forming region properties at the {\it onset} of gas expulsion.  This prompts another key question when quantifying cluster-forming region properties: is an observed embedded cluster caught in the process of turning its gas into stars, or has it just reached the end point of star formation with the intra-cluster gas about to be expelled?
\citet{hig09, hig10} illustrate the problem well.  Their C$^{18}$O mapping of 14 molecular clumps, each forming a star cluster, shows a sequence of star formation efficiencies.  The highest efficiencies are found for clusters associated to C$^{18}$O-emission holes, highlighting that gas dispersal has started there\footnote{An alternative explanation to the C$^{18}$O depletion, however, could be that C$^{18}$O is frozen out on grain surfaces.  Observations of the less-affected nitrogen-containing species NH$_3$ and N$_2$H$^+$ should help clarify the question \citep{ber02} }.  Therefore, an observed \sfe is not enough to assess how the corresponding cluster will respond to gas expulsion.  In Section \ref{ssubsec:loc}, we will suggest an alternative explanation to gas dispersal to explain the C$^{18}$O-emission holes of some cluster-forming clumps.

The characterization of cluster-forming regions has made a significant leap forward thanks to the {\it Spitzer Space Telescope}.  {\it Spitzer} has probed a wide spectrum of stellar densities in star-forming environments of the Solar Neighbourhood -- from relative isolation to star clusters (although the Trapezium region of the Orion Nebula Cluster remains unresolved with that facility).  {\it Spitzer}-surveys have led to a new picture, that is, star clusters are the emerging stellar peaks of wider star-forming regions \citep{all07,eva09,gut11}.  Understanding the physics of cluster formation and the properties of star-forming regions at large are therefore now tightly entwinned topics.  

\citet{gut11} identifies a positive power-law trend between the local surface densities of molecular gas and young stellar objects (YSOs) in eight molecular clouds located within 1\,kpc from the Sun.  On the average, the relation between the local YSO surface density, $\Sigma_{YSO}$, and the local gas surface density, $\Sigma_{g}$, follows $\Sigma_{YSO} \simeq 10^{-3} \Sigma_{g}^{\alpha}$, with $\alpha \simeq 2$ and the surface densities in units of $\Ms \cdot pc^{-2}$.
Their result is based on mapping {\it Spitzer}-identified YSO spatial distributions with a $n^{th}$ nearest neighbour scheme, and tracing the molecular gas by near-infrared extinction.  In most molecular clouds of the sample, the power-law scaling is affected by a large scatter \citep[$\sim 1$\,dex, see fig.~9 in][]{gut11}.  It is tightest in the Ophiuchus and MonR2 clouds where the power-law index $\alpha$ is $1.9$ and $2.7$, respectively.  One of the purposes of this contribution is to show how this star formation law and the physics of cluster formation are related.  

There is increasing evidence for a bimodal cluster formation process.  \citet{dac09} find that the half-light radius distribution of extragalactic old globular clusters is bimodal.  Similarly, \citet{bau10} identify two populations of globular clusters in the outer halo of our Galaxy based on the ratio of their half-mass to Jacobi radii.  They suggest this bimodality to be an imprint of cluster formation, rather than a consequence of 13\,Gyr of dynamical evolution in the Galactic halo \citep[see also ][]{elm08}.  In addition, a bimodal cluster formation agrees with the finding of \citet{pfa09} that, in the Galactic disc, star clusters younger than 20\,Myr unfold along two distinct sequences in the space of cluster volume density versus cluster radius (her fig.~2).  She coins them the {\it starburst} and {\it leaky} clusters.  
[\citet{mai01} identifies a similar dichotomy in the structural properties of a sample of nearby extragalactic clusters.]
\citet{pfa09} also notes that the embeddded clusters of \citet{lad03} define a precursor sequence to the leaky clusters (her fig.~4).  That is, these embedded clusters seem to be the leaky-cluster progenitors, with starburst cluster precursors still to be identified.  

In a follow-up study, \citet{pfa11} shows that the embedded-cluster sequence obeys $\rho_{ecl} \propto r_{ecl}^{-1.3}$, where $\rho_{ecl}$ is the cluster mean volume density (her fig.~2).  This is reminiscent of a relation of constant mean surface density, i.e. $\rho_{ecl} \propto r_{ecl}^{-1}$, as put forward by \citet{ada06} and \citet{all07}.  She presents  the embedded-cluster sequence as a {\it growth} sequence, that is, clusters are still in the process of building their stellar mass and their properties may therefore not be  representative of the conditions at gas expulsion onset.  The embedded-cluster scaling $\rho_{ecl} \propto r_{ecl}^{-1.3}$ equates with $m_{ecl} \propto r_{ecl}^{1.7}$.  This implies that, as the embedded-cluster (stellar) mass $m_{ecl}$ increases with time, so does the radius $r_{ecl}$, while the mean volume density $\rho_{ecl}$ decreases.  In other words, in this scenario, star formation propagates outwardly.  The cluster central regions form first, and outer shells of stars are added with time.  
This is an extremely appealing concept for the following reason.  The timescale relevant for star formation is the free-fall time of the star-forming gas, $\tff$ \citep{elm07,kru07}.  Since $\tff$ scales with the gas volume density as $\tff \propto \rho_g^{-1/2}$ and since molecular clumps have radial density gradients \citep{beu02,mue02,pir09}, the gas free-fall time is  shorter in the denser inner regions of molecular clumps than in their outskirts.  That is, star formation is faster closer to the clump centre than towards the clump edge.  

Here comes a subtle difference, however.  Is star formation in the outer regions of a molecular clump {\it delayed}, or does star formation start at the same time all through the molecular clump, albeit at a {\it slower} rate at larger distance from the clump centre?  The scenario devised by \citet{pfa11} clearly fits the first hypothesis.  In contrast, a molecular clump forming stars on a radially-varying timescale would match the second hypothesis.  

Considerable efforts have been dedicated to the hydrodynamical simulations of star formation in molecular gas \citep[e.g.][]{kle98, bat03, bon08}.  These simulations are computationally expensive and hence limited in terms of the modelled gas mass \citep[e.g. $10^4\,\Ms$, ][]{bon08}.  Parametric semi-analytical studies remain therefore useful, especially  to browse the parameter space extensively.

In this contribution, we build on the concept of \sfe per free-fall time originally introduced by \citet{kru05}, namely, the fraction of an object's gaseous mass that is processed into stars over one free-fall time at the mean density of the object\footnote
{Note that \citet{kru05} use the terminology `star formation {\it rate} per free-fall time', or SFR$_{ff}$ for short.  We prefer to refer to the `star formation {\it efficiency} per free-fall time' and denote it $\epsilon_{ff}$.}.  
We generalize their approach by defining a {\it local} free-fall time.  That is, we present a new model for star-forming regions which hinges on the radially-varying free-fall time of molecular clumps.  We show how it provides an elegant and coherent picture accounting for both the local
\footnote{We refer to the \citet{gut11} star formation law as a {\it local} law to make it distinct from the {\it global} star formation law of \citet{ken98} which builds on galaxy-integrated surface densities.} star formation law $\Sigma_{YSO} \propto \Sigma_{gas}^{2}$ recenly highlighted by \citet{gut11}, and the growth sequence of embedded clusters postulated by \citet{pfa11}.

Our paper is organised as follows.  In Section \ref{sec:mod}, we present the model and show how it explains straightforwardly the scaling law of \citet{gut11}.  In Section \ref{sec:ecl}, we fold the model with a surface density threshold.  This equates with accounting for a contaminating stellar background against which the star-forming region is projected and observed.  We deduce the corresponding $r_{ecl}$-$\rho_{ecl}$ relation, where $r_{ecl}$ and $\rho_{ecl}$ are the radius and mean volume density of the part of the star-forming region which `emerges above' the stellar background.  We find the predicted $r_{ecl}$-$\rho_{ecl}$ relation to be in good agreement with the embedded-cluster sequence of \citet{pfa11}.  Section \ref{sec:conseq} gives an overview of the model consequences in the framework of cluster gas expulsion.  Finally, we outline some future-work directions in Section \ref{sec:fut} and present our conclusions in Section \ref{sec:conc}.

% -----------------------------------------------
\section{Free-fall time driven star formation}
\label{sec:mod}
% ------------------------------------------------
We start by studying how a spherical gas clump with a radial density gradient builds its stellar component as a function of time and space, under the assumption of a constant \sfe {\it per free-fall time}, $\eff$.  Because of the density gradient, the free-fall time must be defined locally, i.e. $\tff(r)$, and is an increasing function of the distance $r$ from the clump centre.  Consequently, star formation proceeds more quickly in the clump central regions than in its outskirts.  This leads to a density profile for the stellar component steeper than the initial density profile of the clump.  

Prior to going into detailed numerical simulations, we present a simple analytical approximation which allows to grasp the relevant physics.

% ..................................................
\subsection{Analytical Insights}
\label{subsec:insight}
% ...................................................
Assuming spherical symmetry, the volume density profile, $\rho_{0}(r)$, of a molecular clump of mass $M_0$, radius $R_0$, and density index $p_0$, obeys:
\begin{equation}
\rho_{0}(r)=k_{\rho, 0} \, r^{-p_0} = \frac{3-p_0}{4 \pi} \frac{M_0}{R_0^{3-p_0}} r^{-p_0} \,,
\label{eq:rho0}
\end{equation}
where $r$ is the distance from the clump centre.  The subscript `$0$' refers to the gas properties prior to star formation (i.e. at $t=0$: the clump is made of gas only).
The factor $k_{\rho, 0}$ comes from integrating the density profile over the clump volume, i.e. $M_0 = \int_0^{R_0} 4 \pi r^2 \rho_0(r)dr$.  The density index of star-forming molecular clumps is observed to range from $\simeq 1.5$ to $\simeq2.0$.  We therefore assume $1.5 \lesssim p_0 \lesssim 2.0$ \citep{beu02,mue02} \footnote{Strictly speaking, the density indices inferred by \citet{beu02} and \citet{mue02} characterize {\it star-forming} molecular clumps.  They may thus be different from $p_0$ which, in this model, is the gas density index {\it prior} to star formation.  We shall come back to this point in Section \ref{ssec:prof}.}.  

In this model, we assume that star-forming molecular clumps experience neither significant outflows or inflows, nor do they contract or expand during star formation.  We also neglect the potential migration of YSOs after their formation.  As a result of these assumptions, at any time $t$ after the onset of star formation, the clump total mass (gas + stars) is preserved, i.e. $M_{\star}(t) + M_g(t) = M_0$, as well as the clump radius $R_0$ and the spatial distribution of the clump mass, i.e. $\rho_{\star}(t,r) + \rho_g(t,r) = \rho_0(r)$.
The subscripts `$g$' and `$\star$' refer to the properties of, respectively, the unprocessed gas and stellar component at any time $t>0$.  As an example, we write $\rho_{g}$ and $\rho_0$ for the {\it current} and {\it initial} gas volume densities.  Finally, we consider that the \sfe per free-fall time, $\eff$, is constant.  
In this class of model, star formation takes place all over the clump -- albeit at a slower rate in the outskirts than in the centre.  The limiting radii of the clump and of its stellar component are therefore equal: $R_0=R_{\star}$.  In what follows, we shall refer both radii as $R$.  \\

Now let us consider that, at any time $t$, a fraction $\eff$ of the gas mass at radius $r$ is turned into stars every local  free-fall time.  Because of the volume density gradient of the molecular clump, star formation proceeds more quickly in its centre than in its outskirts.  As a result, the volume density profile of the stellar component built by the clump, $\rhost(t,r)$,  is {\it steeper} than $\rho_0(r)$.  We insist that this stems from the shorter free-fall time at smaller radius, {\it not} from a higher $\eff$ in the clump centre.  The free-fall time of the gas at radius $r$ and time $t$ obeys: 
\begin{equation}
\tau_{ff}(t,r)=\sqrt{\frac{3\pi}{32G\rho_{g}(t,r)}},
\label{eq:tff}
\end{equation}
with $G$ the gravitational constant and $\rho_{g}(t,r)$ the volume density profile of the unprocessed gas at time $t$.

\begin{figure}
\includegraphics[width=80mm]{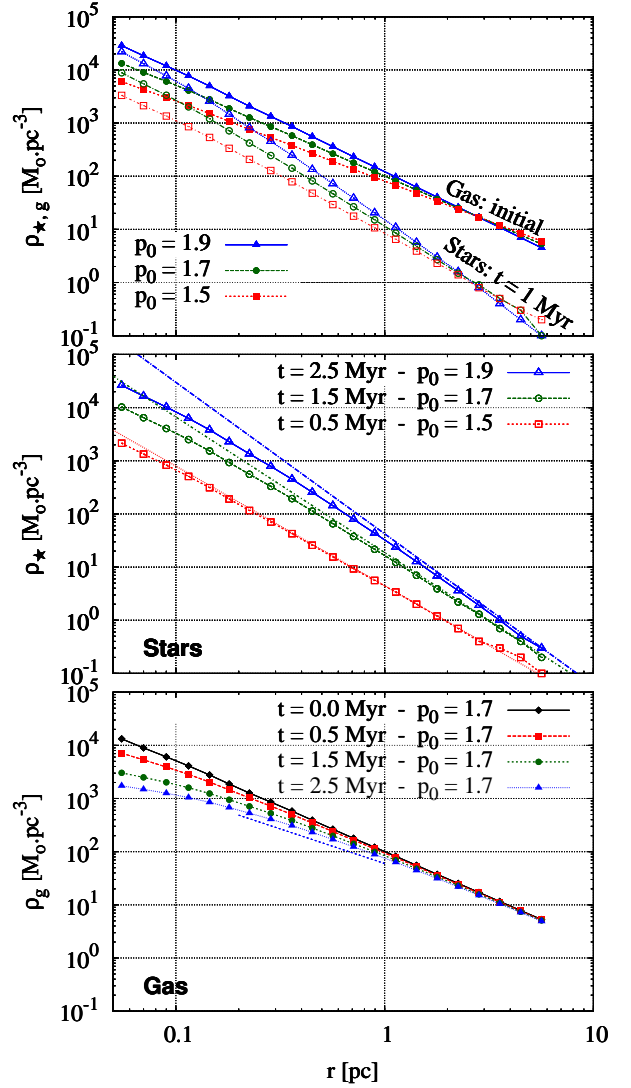}
\caption{{\it Top panel:} Initial gas volume density profiles (filled symbols) and their associated steeper stellar density profiles (open symbols).  The latter are obtained from the numerical model one million years after star formation onset, i.e. $t = 1\,Myr$, for three distinct $p_0$ density indices (see key).  The spherical molecular clump has a mass $M_0 \simeq 10^4\,\Ms$, a radius $R \simeq 6\,pc$, and the \sfe per free-fall time is $\eff = 0.1$.  {\it Middle panel:} Stellar density profiles from the numerical model (lines with open symbols) for the times $t$ and density indices $p_0$ quoted in the key.  Each symbol-free line is the corresponding analytical upper limit (Eq.~\ref{eq:uplim}).  {\it Bottom panel:} Time evolution of the gas volume density profile for $p_0=1.7$.  The star-formation-driven depletion of the gas in the central region of the molecular clump -- where the free-fall time is the shortest -- is clearly highlighted.  The symbol-free dotted line over the range 0.2-1\,pc has a slope of $-1.3$, i.e. shallower than the initial slope $-1.7$.  \label{fig:rhoprof} }
\end{figure}

Let us assume that the volume density profile of the stellar component is a power law too, with a constant density index $q$:
\begin{equation}
\rhost(t,r) = k_{\rho, \star}(t) \, r^{-q} = \frac{3-q}{4\pi} \frac{M_{\star}(t)}{R^{(3-q)}} \, r^{-q}\;. 
\label{eq:rhost}
\end{equation}
As we shall see from the numerical modelling, this is a realistic approximation (see top and middle panels of Fig.~\ref{fig:rhoprof} and top panel of Fig.~\ref{fig:vh}).  
In Eq.~\ref{eq:rhost}, $M_{\star}(t)$ is the total stellar mass contained by the clump at time $t$.  As previously, the factor $k_{\rho, \star}$ stems from integrating the density profile over the entire clump volume, i.e. $M_{\star}(t) = \int_0^{R} 4 \pi r^2 \rho_{\star}(t,r)dr$.  

Because the stellar component is steeper than the initial gas density profile, we have $q>p_0$.  How much steeper is $q$ compared to $p_0$?
Let us assume for a moment that, at any time $t$ and radius $r$, the gas mass keeps dominating the stellar mass.  That is, in $\rho\st(t,r)+\rho_g(t,r)=\rho_0(r)$, we assume $\rho\st(t,r) << \rho_g(t,r)$, which leads to $\rho_g(t,r) \simeq \rho_{0}(r)$, i.e. the gas density profile does not evolve significantly.  Therefore, the local free-fall time (Eq.~\ref{eq:tff}) is approximately constant too:
\begin{equation}
\tau_{ff}(t,r) \simeq \tau_{ff}(t=0,r)\;. 
\label{eq:steady_tff}
\end{equation}
Now consider a shell of thickness $dr$ at a distance $r$ from the clump centre and of initial gas mass $dm_0(r) = 4 \pi  r^2 \rho_0(r) dr$.  Under the assumption of a constant free-fall time (Eq.~\ref{eq:steady_tff}), the stellar mass formed by this shell at time $t$, $dm\st(t,r)$, follows from:
\begin{equation}
dm\st(t,r) \lesssim \eff \frac{t}{\tff(t=0,r)} dm_0(r),
\label{eq:mstap}
\end{equation}
where $t/\tau_{ff}(t=0,r)$ is the number of free-fall times elapsed since star formation started.
The rhs of Eq.~\ref{eq:mstap} defines an {\it upper limit} to the actual mass in YSOs in the shell at time $t$, for two reasons.  Firstly, as time goes by, the \sfe per free-fall time is applied to an ever lower gas mass, i.e. $dm_g(t,r) < dm_0(r)$. Secondly, the steady decrease of the gas mass lengthens the local free-fall time, i.e. we have ${\tff(t,r)} \geq {\tff(t=0,r)}$ instead of Eq.~\ref{eq:steady_tff}.  Equation \ref{eq:mstap} can be rewritten:
\begin{equation}
\rho\st(t,r) \lesssim \eff \frac{t}{\tff(t=0,r)} \rho_0(r),
\label{eq:rhostap}
\end{equation}
Combining Eqs.~\ref{eq:tff} and \ref{eq:rhostap} leads to:
\begin{equation}
\rhost(t,r) \lesssim \sqrt{\frac{32 G}{3 \pi}} \cdot \eff  \cdot t  \cdot \left[\rho_0(r)\right]^{3/2}\;.
\label{eq:uplim0}
\end{equation}
Introducing Eq.~\ref{eq:rho0} then gives:
\begin{equation}
\rhost(t,r) \lesssim \sqrt{\frac{32 G}{3 \pi}} \cdot \eff \cdot t \cdot k_{\rho_0}^{3/2} \cdot r^{-3p_0/2}\;.
\label{eq:uplim}
\end{equation}

The rhs of Eq.~\ref{eq:uplim} is shown as the symbol-free lines in the middle panel of Fig.~\ref{fig:rhoprof}.  The lines with open symbols show the exact solutions obtained either numerically (Section \ref{subsec:num}) or analytically (Section \ref{subsec:anly}) and which account for the time variations of the gas mass and of the local free-fall time.  The panel highlights clearly that the analytical approximation provides an upper limit to the actual stellar density profile.  We assume $M_0 \simeq 10^4\,\Ms$, $R \simeq 6\,pc$ and $\eff = 0.1$ (values discussed in Section \ref{subsec:num}).  Star-formation durations $t$ and density indices $p_0$ are given in the key.  

The rhs of Eq.~\ref{eq:uplim} works best as an analytical approximation for $\rhost(t,r)$ in the clump outskirts where the local free-fall time is the longest.  That is, the hypothesis $\rho_g(t,r) \simeq \rho_0(r)$ remains valid for the time-span of our simulations (up to 2.5\,Myr) and Eq.~\ref{eq:uplim} can be seen as an equality (rather than an upper limit).  In the inner regions, however, the shorter free-fall time implies a gas depletion quicker than in the outskirts.  This eventually results in $\rho_g(t,r) < \rho_0(r)$, ${\tff(t,r)} > {\tff(t=0,r)}$ (star formation slows down), and the rhs of Eq.~\ref{eq:uplim} gives a firm upper limit.  
This differential behaviour between the inner and outer regions means that the actual stellar density profile is shallower than given by the rhs of Eq.~\ref{eq:uplim} (compare the symbol-free lines and the lines with open symbols in the middle panel of Fig.~\ref{fig:rhoprof}, especially for the $p_0=1.9$ model).  In other words, the stellar density profile is steeper than the initial gas density profile (Eq.~\ref{eq:uplim0}) by {\it at most} a factor 1.5, e.g. an initial gas density index $p_0 = 1.7$ gives rise to a star density index $q \lesssim 3p_0/2 = 2.55$.  \\

We now have to convert the volume density profiles into (projected) surface density profiles, since observer-retrieved quantities are surface densities.  For power-law volume density profiles, the surface density profiles are shallower than their volume counterparts by 1\,dex.  Therefore, the surface density profiles of the gas, initially and at time $t$, and of the stars obey:
\begin{equation}
\Sigma_{0}(s)=k_{\Sigma, 0} \, s^{-p_0+1} = \frac{3-p_0}{2 \pi} \frac{M_0}{R^{3-p_0}} \, s^{-p_0+1} \,,
\label{eq:sig0}
\end{equation}

\begin{equation}
\Sigma_{g}(t,s)=k_{\Sigma, g} \, s^{-p+1} = \frac{3-p}{2 \pi} \frac{M_g(t)}{R^{3-p}} \, s^{-p+1} \,,
\label{eq:sigg}
\end{equation}

and 

\begin{equation}
\Sigst(t,s) = k_{\Sigma, \star} \, s^{-q + 1} = \frac{3-q}{2 \pi} \frac{M_{\star}(t)}{R^{3-q}} \, s^{-q+1}\,,
\label{eq:sigst}
\end{equation}
respectively.
The factors $k_{\Sigma, (0,g,\star)}$ come from integrating the surface densities over the whole molecular clump surface: $M_{(0,g,\star)} = \int_0^{R} \Sigma_{(0,g,\star)}(s) \cdot 2 \pi s \cdot ds$, where $s$ is the projected distance from the clump centre.  That is, $s$ is a two-dimensional distance on the plane of the sky, while $r$ is a three-dimensional distance.

With the surface density profiles (Eqs.~\ref{eq:sig0}, \ref{eq:sigg} and \ref{eq:sigst}) derived above and an estimate of the star density index ($q \lesssim 3p_0/2$), we are now ready to infer the local star formation law $\Sigst(t,s) \propto \Sigma_g(t,s)^{\alpha}$ predicted by the analytical approximation.
Eliminating the two-dimensional distance $s$ between Eqs.~\ref{eq:sigg} and \ref{eq:sigst}, we obtain:
\begin{equation}
\Sigst(t,s) = \frac{(3-q)M\st(t)}{[(3-p)M_g(t)]^{\frac{q-1}{p-1}}} \cdot (2\pi R^2)^{\frac{q-p}{p-1}}\cdot \left( \Sigma_g(t,s) \right)^{\frac{q-1}{p-1}}\;.
%\frac{ \frac{3-q}{2\pi} \frac{M\st (t)}{R^{3-q}} }{ \left( \frac{3-p}{2\pi} \frac{M_g(t)}{R^{3-p}} \right)^{\frac{q-1}{p-1}}} 
\label{eq:sfl_mod0}
\end{equation}

Under our assumption that the gas mass dominates the stellar mass all through the clump, 
$\Sigma_0(s) = \Sigma_g(t,s) + \Sigma\st(t,s) \simeq \Sigma_g(t,s)$ and we substitute $\Sigma_{0}(s)$ to $\Sigma_{g}(t,s)$ in Eq.~\ref{eq:sfl_mod0}.  This leads to:
\begin{equation}
\Sigst(t,s) = \frac{(3-q)M\st(t)}{[(3-p_0)M_0]^{\frac{q-1}{p_0-1}}} \cdot (2\pi R^2)^{\frac{q-p_0}{p_0-1}} \cdot \left( \Sigma_0(s) \right)^{\frac{q-1}{p_0-1}}\;.
%\frac{ \frac{3-q}{2\pi} \frac{M\st (t)}{R^{3-q}} }{ \left( \frac{3-p_0}{2\pi} \frac{M_0}{R^{3-p_0}} \right)^{\frac{q-1}{p_0-1}}} \cdot \left( \Sigma_0(s) \right)^{\frac{q-1}{p_0-1}}\;.
\label{eq:sfl_mod}
\end{equation}
Considering the index $\alpha$ of this relation
\begin{equation}
\alpha = \frac{q-1}{p-1} \simeq \frac{q-1}{p_0-1},
\label{eq:sfl_index}
\end{equation}
it immediately appears that a density profile steeper for the stellar component than for the gas (i.e. $q>p_0$) is conducive to $\alpha >1$, as found by \citet{gut11}.  Were the density indices for the gas and stars identical (i.e. $q = p_0$, equivalent to no radial variations of the local star formation efficiency; see also Section \ref{ssubsec:loc}), $\alpha$ would be unity.  There is therefore a causal link between 
{\it (i)} the slope of the relation between the gas and YSO surface densities (Eq.~\ref{eq:sfl_mod}), and  
{\it (ii)} how much steeper the volume density profile of the stars is compared to that of the gas (Eq.~\ref{eq:uplim0}).
Adopting $p_0 = 1.7$ and $q \lesssim 3p_0/2 = 2.55$ leads to $\alpha \lesssim 2.2$, in fair agreement with the mean power-law trend observed by \citet{gut11}.  This quantitative result provides us with a strong incentive to refine the $(\Sigst, \Sigma_{g})$ relation by means of numerical simulations.

% ..................................................
\subsection{Numerical Model}
\label{subsec:num}
% ...................................................
In the numerical model, we discretize the spherical molecular clump into successive shells defined by their three-dimensional radius $r$, local volume density $\rho_{g}(t,r)$ and local free-fall time $\tff(t,r)$ at time $t$.  Every instantaneous local free-fall time, a fraction $\eff$ of the gas mass is removed and added to the stellar content:
\begin{equation}
dm_g(t_i,r) = dm_g(t_{i-1},r) - \eff \frac{(t_i - t_{i-1})}{\tff(t_{i-1},r)} \cdot dm_g(t_{i-1},r).
\label{eq:mgnum}
\end{equation}
\begin{equation}
dm\st(t_i,r) = dm\st(t_{i-1},r) + \eff \frac{(t_i - t_{i-1})}{\tff(t_{i-1},r)} \cdot dm_g(t_{i-1},r).
\label{eq:mstnum}
\end{equation}
The local gas free-fall time (Eq.~\ref{eq:tff}) is then updated.  As the gas gets depleted, star formation slows down, an effect unaccounted for in the analytical approximation of Section \ref{subsec:insight}.  

The numerical results are shown in Fig.\ref{fig:rhoprof}, for the case of a spherical gas clump of mass $M_0 = 10^4\,\Ms$, radius $R = 6$\,pc and \sfe per free-fall time $\eff = 0.1$.  Our choice of the mass and radius stems from inspecting the extinction map of the MonR2 molecular cloud and its spatial distribution of {\it Spitzer}-identified YSOs \citep[][their fig.~1]{gut11}.  We focus on the largest concentration of YSOs (at a right ascension of $\simeq 06^h08^m$ and declination of $\simeq -6^{\circ} 24'$) for which we adopt a radius $R = 6$\,pc so as to include the lowest YSO surface densities.  We adopt $M_0 = 10^4\,\Ms$ as a rough estimate of the initial gas mass within $R = 6$\,pc given that this region represents a significant fraction of the MonR2 cloud whose total mass is $25,800\,\Ms$ \citep[table~1 in][]{gut11}.  We plan to map our model onto detailed observational data in the near future. In the mean time, we stress that the estimate of $\eff$ we adopt to make our model match the averaged local star formation law, $\Sigma\st = 10^{-3} \Sigma_g^2$, depends on the chosen values for $M_0$ and $R$ (see below and top panel of Fig.~\ref{fig:sfl}).  

As for the upper limit on the time $t$ elapsed since star formation onset, we note that \citet{gut11} study encompasses mostly Class I protostars and Class II pre-main-sequence stars.  Given that the respective average lifetimes of the Class I and Class II phases are $\simeq 0.5$\,Myr and $\simeq 2$\,Myr \citep{eva09}, we adopt 2.5\,Myr as the time-span of our simulations.

The top panel of Fig.\ref{fig:rhoprof} illustrates the initial volume density profile of the gas for different density indices: $p_0 = 1.5, 1.7$ and $1.9$ (plain symbols: red squares, green circles and blue triangles, respectively).  Also depicted are the volume density profiles of the built-in stellar component at $t=1$\,Myr (open symbols with identical symbol/colour-coding), thereby highlighting the steepening of the stellar density profile compared to the initial spatial distribution of the gas.  The gas-to-star steepening equates with a local \sfe higher in the clump central region than in its outskirts.  We come back to this point in Section \ref{ssubsec:loc}.  The middle panel shows the stellar density profiles for the gas density indices $p_0$ and times $t$ quoted in the key.  Each profile is shown along with its upper limit predicted by Eq.~\ref{eq:uplim}.  The difference between the analytical approximation (symbol-free lines) and the numerical model (open symbols) is stronger when the free-fall time is short, e.g. in the clump inner regions, especially for a steep density profile (high $p_0$).  The difference also gets higher for longer star-formation durations $t$.  Nevertheless, the comparison between both demonstrates the excellence of Eq.~\ref{eq:uplim} in providing a back-of-the-envelope estimate of the stellar density profile.
The bottom panel of Fig.\ref{fig:rhoprof} depicts the time evolution of the gas density profile when $p_0=1.7$. Its flattening in the clump centre contrasts markedly with the absence of evolution at the clump edge where the initial free-fall time is the longest.  The instantaneous gas density index, $p$, thus becomes smaller/shallower as $t$ increases.  

% ..................................................
\subsection{Analytical solution}
\label{subsec:anly}
% ...................................................

The actual time-evolution of the star and gas density profiles can also be obtained analytically.   Equations~\ref{eq:mgnum}-\ref{eq:mstnum} correspond to separable first order differential equations.  Using Eq.~\ref{eq:tff}, they can be rewritten as :

\begin{equation}
\frac{\partial \rho_g(t,r)}{\partial t} = -\frac{\eff}{\tff(t,r)} \rho_g(t,r)=-\sqrt{\frac{32G}{3\pi}} \cdot \eff \cdot  \rho_g(t,r)^{3/2}
\label{eq:mgdif}
\end{equation}

\begin{equation}
\frac{\partial \rhost(t,r)}{\partial t} = \frac{\eff}{\tff(t,r)} \rho_g(t,r)=\sqrt{\frac{32G}{3\pi}} \cdot \eff \cdot  \rho_g(t,r)^{3/2}\,,
\label{eq:mstdif}
\end{equation}

and their solutions are:
\begin{equation}
\rho_g(t,r) = \left( \rho_0(r)^{-1/2} + \sqrt{\frac{8G}{3\pi}} \cdot \eff \cdot t  \right)^{-2}\,,
\label{eq:mgsol}
\end{equation}

\begin{equation}
\rhost(t,r) = \rho_0(r) - \left( \rho_0(r)^{-1/2} + \sqrt{\frac{8G}{3\pi}} \cdot \eff \cdot t   \right)^{-2}
\label{eq:mstsol}
\end{equation}

where the clump density profile $\rho_0(r)$ is given by Eq.~\ref{eq:rho0}.  Equations \ref{eq:mgsol} and \ref{eq:mstsol} respectively correspond to the lines with plain symbols in bottom panel of Fig.~\ref{fig:rhoprof} and to the lines with open symbols in top and middle panels of Fig.~\ref{fig:rhoprof} obtained numerically above.  \\

Before going any further, we remind that our model builds on the {\it local} free-fall time, that is, it quantifies the local ({\it at} the radius $r$) collapse of the gas.  The {\it global} collapse of the star-forming molecular clump (i.e. its collapse to its centre) may also be relevant to its evolution.  The free-fall time to the clump centre depends on the mean volume density enclosed {\it within} the radius $r$, $\overline{\rho_0(<r)}$, and may therefore be shorter than the local free-fall time, especially for steep density profiles \citep[e.g. the outskirts of some molecular clumps; see fig.~2 in][]{beu02}.  That is, were the clump evolution driven by gravity only, the gas would collapse to the clump centre faster than it does locally.  The significance of the global collapse with respect to the local one depends therefore on how much supported the gas is by the combined effects of turbulence, magnetic fields, ...   A study of that aspect is beyond the scope of the present paper and we assume for now that the star-formation history of the molecular clump is solely driven by the local collapse of its gas.  
As already quoted in Section \ref{subsec:insight}, once the YSOs have formed at radius $r$ through the gas local collapse, we neglect their potential migration towards other regions of the molecular clump.  The actual behaviour of newly-formed stars depends on their velocities at formation.  If their initial velocity is sub-virial, once decoupled from the gas, they will fall to the cluster centre where they will revirialize \citep[see e.g.][]{gir12a}.

% ..................................................
\subsection{Surface densities and the local star formation law}
\label{subsec:sfl}
% ...................................................

\begin{figure}
\includegraphics[width=80mm]{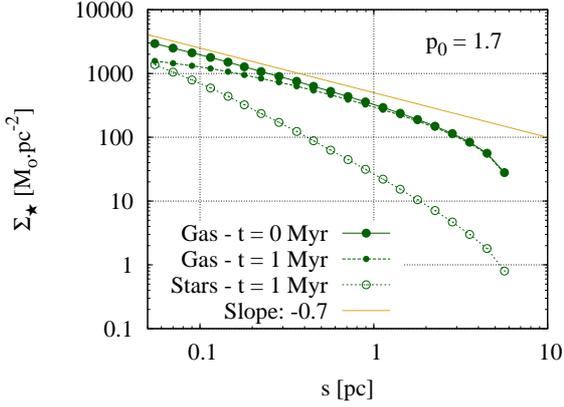}
\caption{Surface density profiles from the numerical model.  Lines with circles from top to bottom: the molecular clump gas initially, the unprocessed gas and the stellar component at $t=1$\,Myr (see key).  The molecular clump is the same as in Fig.~\ref{fig:rhoprof}, and the initial gas density index is $p_0=1.7$.  The solid symbol-free line depicts a slope of $-0.7$, that is, the slope expected for the initial surface density profile of the gas when $p_0=1.7$.  We note that in the clump outskirts, the actual surface density profile is becoming increasingly steeper than the power-law approximation.        \label{fig:sigprof} }
\end{figure}

Following the analytical aproximations obtained in Section \ref{subsec:insight} (Eqs \ref{eq:sigg}-\ref{eq:sigst}), we now derive the numerical solutions for the gas and star {\it surface} density profiles.  To derive the exact surface density profiles, we build on eq.~6 of \citet{par11c} which integrates a power-law volume density profile into the projected mass enclosed within an aperture of radius $s_{ap}$ (see also their fig.~3).  This provides straightforwardly the projected mass within a circular corona of radius $s_{ap}$ and thickness $ds_{ap}$.  Gas and star surface density profiles are shown in Fig.~\ref{fig:sigprof} for the same molecular clump as previously, $p_0=1.7$ and the times $t$ quoted in the key.  The local star-formation law predicted by our numerical model can now be derived and compared with the scaling observed by \citet{gut11}.

\begin{figure}
\includegraphics[width=80mm]{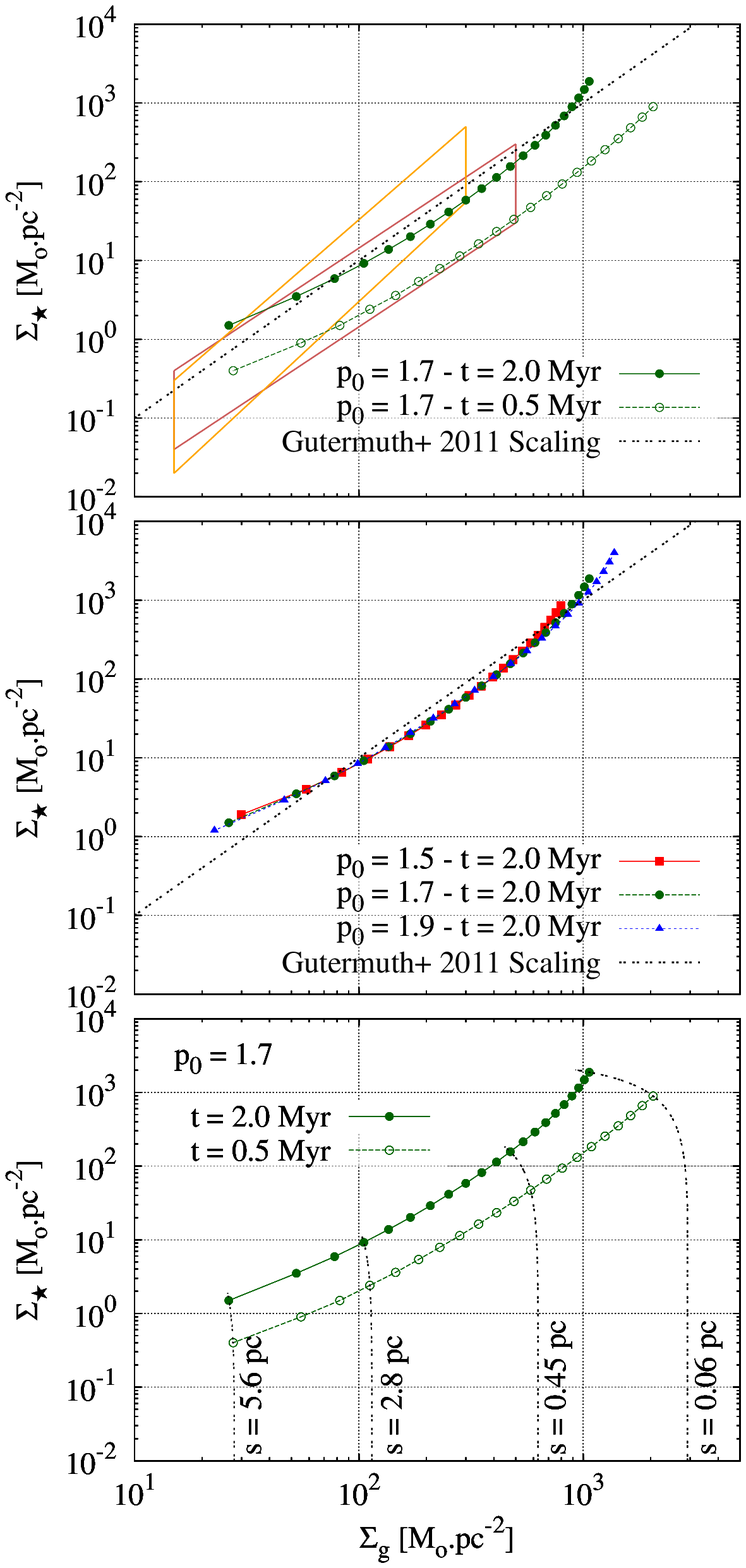}
\caption{Local surface density of YSOs, $\Sigma_{\star}$, in dependence of the local surface density of the unprocessed/observed gas, $\Sigma_{g}$.  {\it Top panel:}  Models for a molecular clump of mass $M_0 = 10^4\,\Ms$, radius $R = 6\,$pc, density index $p_0 = 1.7$, a \sfe per free-fall time $\eff = 0.1$ and the times $t$ quoted in the key.  The dotted (black) line depicts the average star-formation law inferred by \citet{gut11}.  The orange and brown polygons illustrate the associated scatter observed for the MonR2 and Ophiuchus molecular clouds, where it is the smallest.  {\it Middle panel:} Models for a time $t=2$\,Myr and three distinct density indices of the molecular clump, $p_0=1.5$, $1.7$ and $1.9$.  The normalization of the model hardly depends on $p_0$.  {\it Bottom panel:} Same as top panel but completed with the time evolution of the gas and YSO surface densities at four projected distances $s$ from the clump centre ($s$-labelled dotted black lines)  \label{fig:sfl} }
\end{figure}

Figure \ref{fig:sfl} depicts the local surface density of YSOs, $\Sigst$, in dependence of the local surface density of the observed/unprocessed gas, $\Sigma_{g}$.  Both surface densities are in units of $\Ms\cdot pc^{-2}$.  The (black) dotted line in the top and middle panels is the scaling law of \citet{gut11}, i.e. $\Sigst = 10^{-3} \Sigma_{g}^2$.  The top panel shows the numerical model for $p_0=1.7$ at $t=0.5$\,Myr and $t=2.0$\,Myr (lines with open and plain symbols, respectively). 
The agreement between the model at $t=2.0$\,Myr and the mean observed scaling is excellent, both in terms of slope and normalization.  This agreement is parameter-dependent, however, as a molecular clump more massive (hence denser), or a longer time-span, or a higher \sfe per free-fall time, all would shift the normalization upwards.  The $t=2$\,Myr model bends upwards at high surface density.  This is due to the flattening of the gas density profile in the clump centre, that is, $p$ decreases which in turn increases the index $\alpha$ of the local star formation law (Eq.~\ref{eq:sfl_index}; see also bottom panel of Fig.~\ref{fig:sfl}). 

It is worth keeping in mind that $\Sigst = 10^{-3} \Sigma_{g}^2$ defines an {\it average} relation.  The YSO-vs-gas relation observed for each molecular cloud exhibits a considerable scatter \citep[see the panels of fig.~9 in][]{gut11}.  This is because a molecular cloud consists of  several molecular clumps each with its own mass, radius, density index and duration of the star-formation process.  
Therefore, a molecular cloud combines several correlations, each with its own normalization, corresponding to the several molecular clumps it contains.  This leads to a trend in the observed ($\Sigma_g, \Sigma_{\star}$) space, rather than the one-to-one relation predicted for a single molecular clump and depicted in Fig.~\ref{fig:sfl}.
Other reasons for the spread in the observed ($\Sigma_g, \Sigma_{\star}$) relation include: YSO migration and gas dispersal via stellar feedback processes \citep{gut11}, non-spherical molecular clumps.  \citet{gut11} note the scatter to be the smallest in the MonR2 and Ophiuchus molecular clouds, which they represent as the green and blue parallelograms in their fig.~9.  They are reproduced as the orange and brown parallelograms in the top panel of our Fig.~\ref{fig:sfl}.

The middle panel of Fig.~\ref{fig:sfl} illustrates how the model responds to varying the gas density index $p_0$.  For steeper profiles, the clump inner regions become denser at the expense of the outskirts. This stretches the model towards both lower and higher surface densities, although the normalization is hardly affected.  

The bottom panel completes the top panel with the time evolution of the gas and star surface densities at four projected distances $s$ from the clump centre (black dotted lines: central regions are to the right, outskirts are to the left).  As long as a few per cent only of the gas have been turned into stars, a track evolves vertically as the gas surface density does not decrease significantly.  Following its vertical leg, however, the track bends leftwards, thereby embodying the combined stellar mass increase and gas mass decrease.  
We note that a similar plot is provided in fig.~13 of \citet{gut11}.  They assume a star formation law where the star formation rate per unit area shows a power-law dependence on the gas column density.  It thus differs from our model which builds on the time-evolution of volume densities.

% ..................................................
\subsection{Distribution of YSO surface densities}
\label{subsec:Sigdist}
% ...................................................
\begin{figure}
\includegraphics[width=80mm]{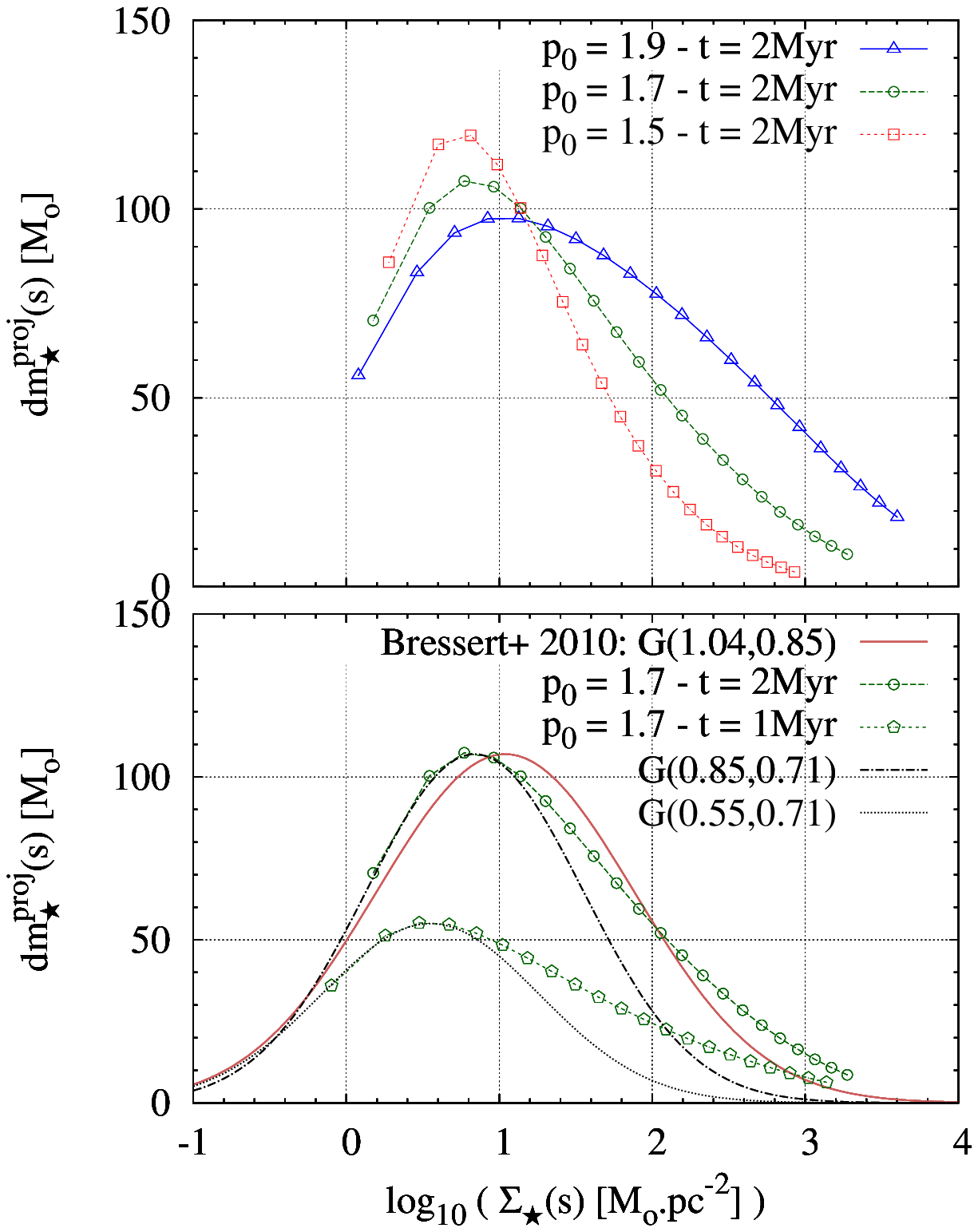}
\caption{{\it Top panel: } Distribution of stellar surface densities predicted by our model for a clump of mass $M_0 \simeq 10^4\,\Ms$,  radius $R \simeq 6\,pc$, and \sfe per free-fall time $\eff = 0.1$ (clump density index $p_0$ and time $t$ since star formation onset are given in the key).  {\it Bottom panel: } Two $p_0=1.7$ models are shown at time $t=1$\,Myr and $t=2$\,Myr for the same molecular clump as above (lines with open symbols).  The dotted and dash-dotted (black) lines depict two Gaussians fitting their low density regime.  The solid (red) line is the Gaussian given by \citet{bre10} to describe the observed local surface density distribution of YSOs in the Solar Neighbourhood.  Mean and standard deviation of all three Gaussians are given in brackets.      \label{fig:Sigdist} }
\end{figure}

\citet{bre10} recently obtained the distribution of local surface densities of {\it Spitzer}-detected YSOs in star-forming regions of the Solar Neighbourhood (distance smaller than 500\,pc).   Their observed distribution is well-approximated by a Gaussian function with a peak at $\simeq 22$\,YSOs pc$^{-2}$ and a standard deviation of 0.85 in $\log_{10}\Sigma_{YSOs}$ (see their fig.~1).  A fully consistent comparison with the result of \citet{bre10} still requires our model to be extended to entire molecular clouds (i.e. to cumulate many molecular clumps) since their observed distribution encompasses a dozen of local star-forming regions.  
It is nevertheless interesting to see what stellar surface density distribution our model predicts at its current stage of development.  

Building on the stellar surface density profiles derived in Section \ref{subsec:sfl}, Fig.~\ref{fig:Sigdist} shows the distribution of the logarithmic stellar surface densities, that is, the projected mass of stars enclosed within an annulus of radius $s$, $dm\st^{proj}(t,s)$, as a function of the annulus logarithmic surface density, $\log_{10} (\Sigst(t,s))$.  Units are $\Ms$ and $\Ms \cdot pc^{-2}$.  The mass, radius and star formation efficiency per free-fall time of the molecular clump are as previously.  The lines with open symbols are the model predictions for the clump density indices $p_0$ and times $t$ given in the key.  In the bottom panel, two models are depicted for $p_0=1.7$: $t=1$\,Myr and $t=2$\,Myr (green curves with open pentagons and circles, respectively).  The low surface-density regime of each model is fitted by a Gaussian (black dotted/dash-dotted lines), with mean $\log_{10}\Sigst=0.55$ ($t=1$\,Myr) and $\log_{10}\Sigst=0.85$ ($t=2$\,Myr), and a standard deviation of 0.71.  They illustrate the growth of the stellar mass and stellar surface density as time goes by.  All models are bell-shaped, with wider distributions for steeper density profiles, as expected.  This suggests that the shape of the observed distribution could be used as a probe into the density profile of star cluster parent clumps.  

The Gaussian with which \citet{bre10} describe their data is shown as the bottom panel solid (red) line.  Assuming a mean stellar mass of $0.5\Ms$ per YSO, the surface density at their peak ($\simeq 22$\,YSOs pc$^{-2}$)  equates with $\log_{10}\Sigst=1.04$.  Their standard deviation is 0.85 and our normalization is arbitrary.
At $t=2$\,Myr, our $p_0=1.7$ model is in good agreement with the distribution observed in the Solar Neighbourhood.  
In particular, the observed distribution agrees with our model prediction better than with the surface density distribution of sink particles of the hydrodynamics simulations of \citet{bon08} \citep[see fig.~2 in][]{kru12}. 
It remains to be seen how including heavily crowded regions in the \citet{bre10} 's sample would affect the comparison.  Access to high-surface density regions (e.g. the core of the Orion Nebula Cluster) -- which {\it Spitzer} fails to resolve and which therefore affects the high-density tail of the observed distribution -- would actually allow us to test our model more extensively.  
Finally, we emphasize that the observation of a smooth distribution in $\log_{10}\Sigst$ alone does not allow one to conclude that star formation is not made of multiple discrete modes \citep{pfa12}.  For a set of low-mass gas-poor clusters, \citet{gie12} propose that the surface density at the peak of the observed distribution is driven by the degree of early cluster expansion.

In this section, we have developed a star-forming region model building on a constant \sfe per free-fall time in a spherical molecular clump with a radial volume density profile $\rho_0 \propto r^{-p_0}$.  When $p_0 \simeq 1.7$ as observed, the model predicts a local star-formation law $\Sigst \propto \Sigma_{g}^\alpha$ with $\alpha \simeq 2$.  In the next section, we combine the model to a surface density threshold imposed by the stellar background against which the star-forming region is seen projected.  We demonstrate that this leads to the time sequence for embedded-cluster development recently proposed by \citet{pfa11}.

% --------------------------------
\section{The observed growth sequence of embedded clusters}
\label{sec:ecl}
% --------------------------------
At first glance, the model presented in Section \ref{sec:mod} departs from the scenario proposed by \citet{pfa11}.  In the scenario she devised based on the data of \citet{lad03}, an embedded cluster grows outwardly, with outer shells of stars added with time.  That is, star formation is {\it delayed} in the outer regions compared to the inner ones.  In contrast, in the present model, star formation proceeds all through the molecular clump albeit at a rate slower in the outskirts than in the centre.  As quoted in Section~\ref{sec:intro}, however, the relation inferred by \citet{pfa11} between the radius and the mean volume density of embedded clusters, i.e.  $\rho_{ecl} \propto r_{ecl}^{-1.3}$, is reminiscent of one of constant mean surface density, i.e. $\rho_{ecl} \propto r_{ecl}^{-1}$.  This suggests that the data of \citet{lad03} are surface-density limited.  Therefore, prior to comparing our results to the growth sequence of \citet{pfa11}, we include in our model this observational bias.  Note that \citet{all07} made a similar point.

\begin{figure}
\includegraphics[width=80mm]{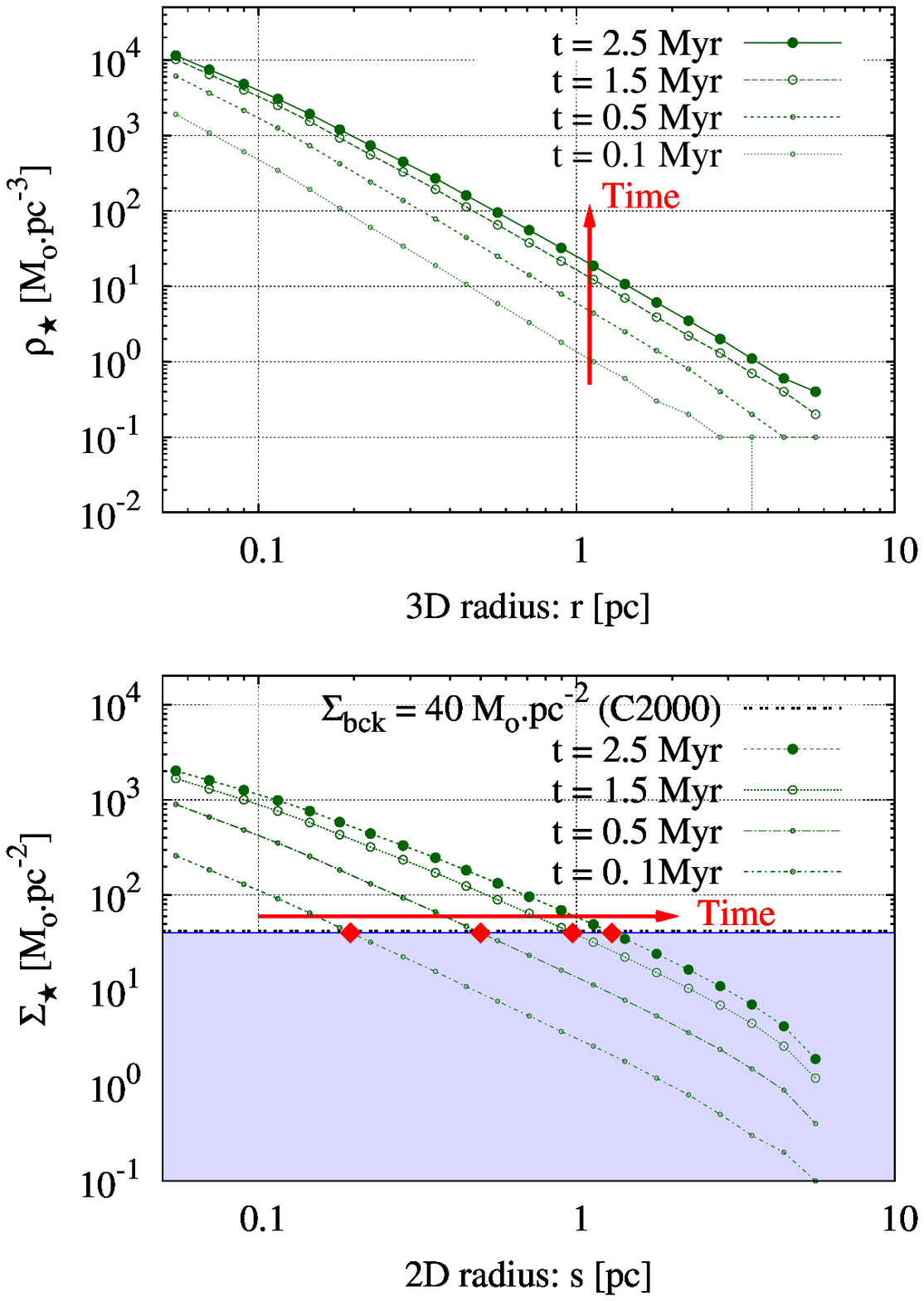}
\caption{{\it Top panel:} Time evolution of the volume density profile of the stellar component for the model of Section \ref{subsec:num}: $M_0 = 10^4\,\Ms$, $R = 6\,$pc, $\eff = 0.1$ along with $p_0 = 1.7$.  Model time-spans $t$ are given in the key. {\it Bottom panel:}  Time evolution of the corresponding surface density profiles superimposed with the surface density limit, $\Sigma_{bck}$, imposed by the stellar background against which the star-forming region is projected \citep[][]{car00}.  The shaded area shows how the surface density threshold conceals the `wings' of the stellar component, leading to observed radii (red diamonds) smaller than the actual one, i.e. $s_{bck} < R$. \label{fig:vh} }
\end{figure}

\begin{figure}
\includegraphics[width=80mm]{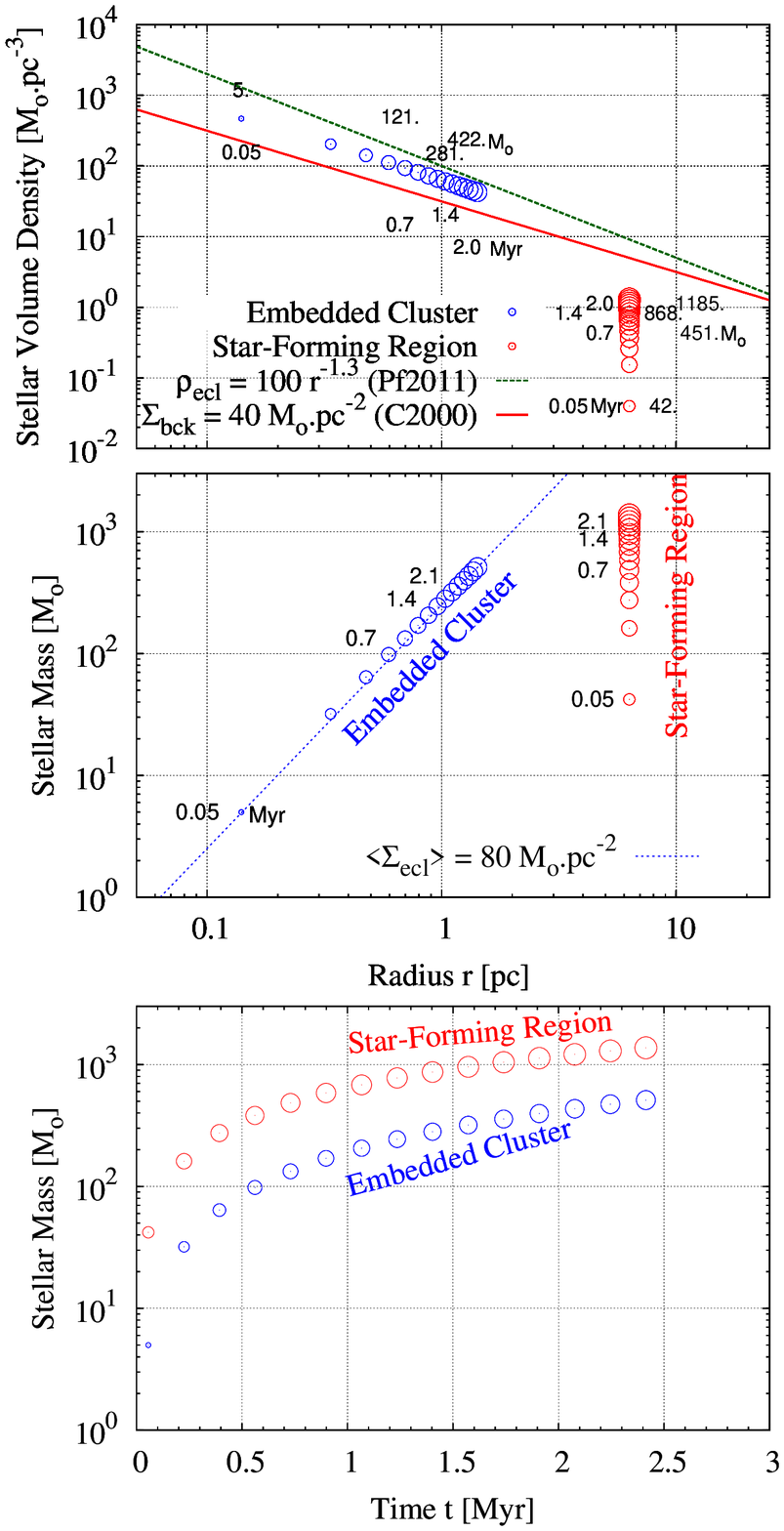}
\caption{{\it Top panel:} Mean volume density in dependence of radius for {\it (i)}~the whole stellar component (vertical sequence of red circles at $R=6$\,pc) and {\it (ii)}~the associated embedded cluster (decreasing sequence of blue circles on the left).  The open circle size scales with the logarithm of the stellar mass.   The embedded cluster refers to the part of the star-forming region seen above the stellar background against which it is seen projected ($\Sigma_{bck} = 40\,M_{\odot} \cdot pc^{-2}$, solid red line).  As a result of the applied surface density cut-off, the predicted embedded-cluster sequence has a constant mean surface density.  Also, it does not differ much from the observed embedded-cluster sequence defined by \citet{pfa11} (dashed green line; see text for details).  The times and stellar masses  are given along the plotted sequences.  {\it Middle panel:} Mass in stars against radius, for the whole star-forming region (red vertical track to the right) and for the developing embedded cluster (blue track to the left).  Quoted to the left of the sequences are the times elapsed since the onset of star-formation (in Myr).  {\it Bottom panel:} Same as the middle panel but for the mass-vs-time space.     \label{fig:eclseq} }
\end{figure}

The top and bottom panels of Fig.~\ref{fig:vh} show the rising with time of the volume- and surface-density profiles of the stellar component for the model of Section \ref{sec:mod} ($M_0 = 10^4\,\Ms$, $R = 6\,$pc, $\eff = 0.1$) with a density index $p_0 = 1.7$.  The bottom panel also includes the surface density cut-off of \citet{car00}: $\Sigma_{bck} = 40\,\Ms \cdot pc^{-2}$ (horizontal dotted line).  The shaded area visualizes the surface density regime with $\Sigma_{\star} < \Sigma_{bck}$.  We note that the embedded clusters of \citet{lad03} are spread over distances from the Sun ranging from $\simeq 100$\,pc to 2.4\,kpc.  Each cluster has therefore its own limiting background surface density.  Since these are unknown, we resort to the single value of $\Sigma_{bck}$ quoted above.  It should be considered as a mean value used for illustrative purposes.  The surface density threshold $\Sigma_{bck} = 40\,\Ms \cdot pc^{-2}$ follows closely the embedded-cluster track defined by \citet[][her fig.~2]{pfa11}.  We will come back to this point in Fig.~\ref{fig:eclseq}. 

The surface density limit imposed by the stellar background conceals the `wings' of the stellar component, whose observed limiting radius is therefore smaller than the actual one, that is, $s_{bck} < R$.  These observed radii, $s_{bck}$, are depicted as (red) diamonds in the bottom panel of Fig.~\ref{fig:vh}.  They define an outward time-sequence, in agreement with \citet{pfa11}.  In other words, while the whole stellar component follows a `vertical' evolution, i.e. the density increases at fixed radius, the fraction of it denser than the stellar background mimics a `horizontal' evolution, i.e. the observed limiting radius increases with time.  In what follows, we refer the part of the stellar component `above' the background as the `embedded-cluster' (non-shaded area in bottom panel of Fig.~\ref{fig:vh}: $\Sigma(s)>\Sigma_{bck}$ when $s<s_{bck}$).  That is, $s_{bck}=r_{ecl}$ with $r_{ecl}$ the embedded-cluster radius.  We define the embedded-cluster mass, $m_{ecl}$, as the stellar mass enclosed within the three-dimensional radius $r_{ecl}$.

For each time-step, we obtain the radius and mass of the embedded cluster and its mean volume density $\rho_{ecl}=3 \cdot m_{ecl}/ (4 \cdot \pi \cdot r_{ecl}^3)$.  
Figure \ref{fig:eclseq} provides three different perspectives of the evolution of the whole stellar component (red open circles) and of its associated embedded cluster (blue open circles): from top to bottom, volume density against radius \citep[parameter space identical to fig.~2 in][]{pfa11}, mass against radius, and mass against time since star-formation onset.  The size of the open circles scales with the logarithm of the stellar mass.

In the top panel, the vertical (red) sequence at a radius $R=6$\,pc depicts the growth with time of the mean volume density of the whole stellar component.  Times in Myr and stellar masses in $\Ms$ are given to the left and right of the sequence, respectively.  
The decreasing (blue) sequence on the left is the predicted embedded-cluster sequence, i.e. $\rho_{ecl}$ versus $r_{ecl}$, with times since star-formation onset and embedded-cluster masses quoted below and above it.  This sequence runs parallel to the surface density threshold $\Sigma_{bck}$ (solid red line) applied to the star-forming region.  The difference in volume density between the threshold $\Sigma_{bck}$ and the predicted sequence ($r_{ecl}$, $\rho_{ecl}$) stems from the density profile of the stellar component, that is, the density {\it at} the limit $r_{ecl} = s_{bck}$ is lower than the mean volume density {\it inside} $r_{ecl}$.
The (green) dashed line depicts the observed embedded-cluster track defined by \citet{pfa11} with which our model (blue open circles) agrees reasonably well.  In a follow-up paper, we will map the parameter space ($M_0$, $R$, $p_0$ and $\eff$) to investigate what patterns emerge in the $r_{ecl}$-$\rho_{ecl}$ space as a function of the input parameters.

The middle panel of Fig.~\ref{fig:eclseq} illustrates the stellar mass growth of the whole star-forming region and of the embedded cluster (red circles at constant radius to the right and blue circles to the left, respectively).  Model time-spans $t$ are  indicated along the sequences.  The bottom panel shows the stellar mass in dependence of the time $t$ since star-formation onset.  As the stellar mass builds up, an ever greater stellar mass fraction makes it above the background limit (see also bottom panel of Fig.~\ref{fig:vh}).  Therefore, the shift between the embedded-cluster and stellar-component sequences tightens.  While the embedded cluster represents about one-tenth of the stellar component total mass at a time   $t=0.1$\,Myr, the cluster mass fraction has risen up to about one-third by $t=2.5$\,Myr.  We therefore conclude that, not only does the stellar mass fraction in clusters depend on the surface density threshold adopted to define clusters  (as opposed to their surrounding `haloes' of stars), it also depends on {\it when the star-formation process started}.  

Here, a remark about the embedded-cluster mass function is worth being made. The mass function of star clusters is a powerful diagnostic tool of their evolution.  For instance, if the mass function slope remains unchanged between the onset and the end of violent relaxation \footnote{
Violent relaxation is the dynamical response of the stellar component to the expulsion of the unprocessed star-forming gas. Its duration -- from  $\simeq 1$\,Myr to several tens of Myr -- depends on the crossing-time of the cluster gaseous precursor and on the observed cluster region.  The denser the gas clump (i.e. the shorter its mean  crossing-time), the smaller the aperture with which the cluster is observed (i.e. the higher the observed cluster density), the faster the cluster evolution  \citep{par12a}}, this implies that cluster infant weight-loss through violent relaxation is mass-independent \citep[e.g.][]{par08b}. 
The embedded clusters of \citet{lad03} have a power-law mass function of slope $\simeq -2$ (their fig.~2), similar to the mass function of young gas-free clusters \citep{cha10}.  We now see that this aspect could be considered under a new viewing angle.   If the clusters compiled by \citet{lad03} mark different stages of their formation process, as \citet{pfa11} suggests, then the slope of the embedded-cluster mass function {\it at gas expulsion} (i.e. when the build-up of the stellar content terminates) may differ from $-2$, since the mass of the clusters at the start of the growth sequence should be corrected for the still missing stellar mass.  This would have consequences as to whether cluster infant weight-loss is mass-independent or not.  

\begin{figure}
\includegraphics[width=80mm]{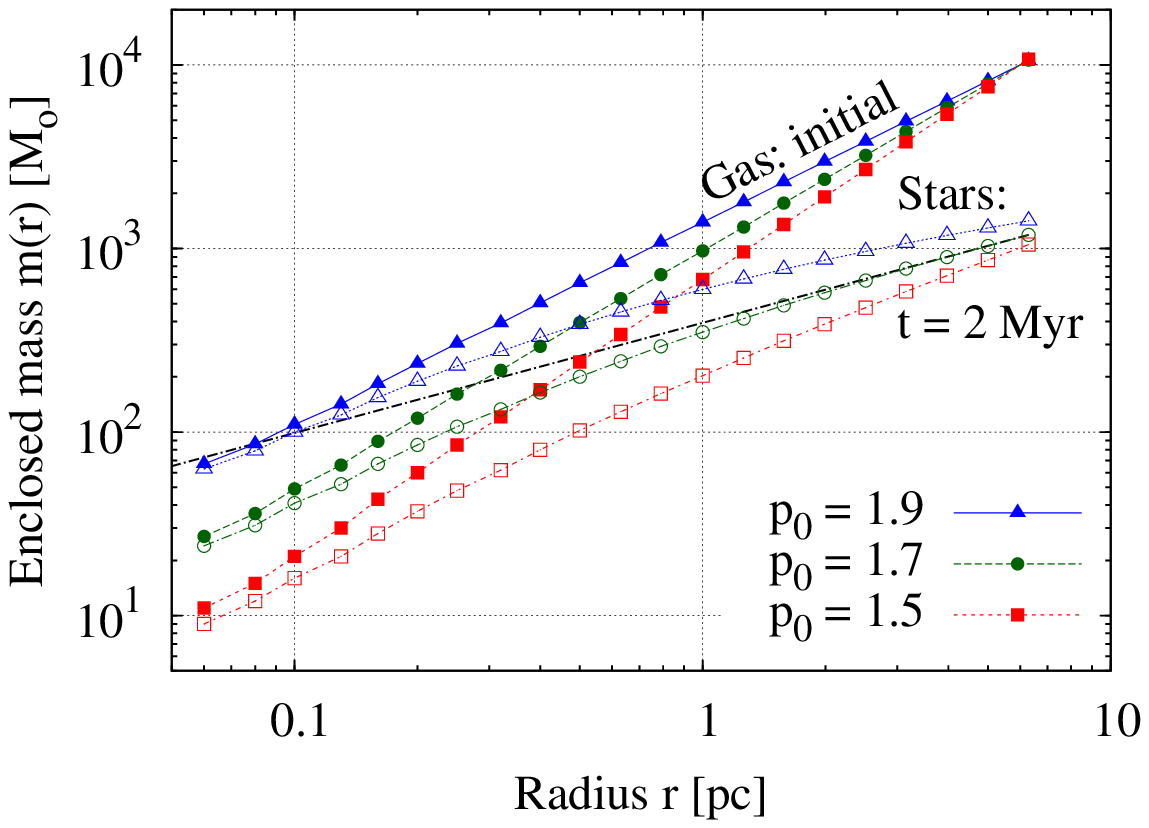}
\caption{Cumulative mass $m(r)$ vs. the three-dimensional radius $r$ for the gas initially (plain symbols) and the stars at $t=2$\,Myr (open symbols) with density indices $p_0$ quoted in the key.  For a power-law density profile of index $p_0$ (Eq.~\ref{eq:rho0}), the slope of the mass distribution in the $r-m(r)$ space is $3-p_0$ (Eq.~\ref{eq:mr}).  The steeper density profiles of the stellar component thus lead to $r-m(r)$ relations shallower than their initial gas analogs (i.e. $3-q$ vs. $3-p_0$).  Note that the slope of the stellar tracks get steeper towards the clump centre due to the shallower stellar density profiles there.  The dash-dotted symbol-free line has a slope of 0.5, as expected for the $r-m(r)$ relation of the stellar component in the low-density regime when $p_0=1.7$ (see text for details) \label{fig:rm3d} }
\end{figure}

For an assumed volume density profile of the stellar component, one can extrapolate the mass of the embedded cluster to that of the whole stellar component by representing the star mass distribution in the $r-m(r)$ space.  
Figure~\ref{fig:rm3d} illustrates the relations between the distance $r$ from the star-forming region centre and the enclosed mass $m(r)$, for the gas initially and the stellar component at $t=2$\,Myr (symbol/colour-coding identical to previously).  For a pure power-law density profile of index $p_0$, radii and masses follow (taking the case of the initial gas density distribution, Eq.~\ref{eq:rho0}):
\begin{equation}
\frac{m_0(r)}{M_0} = \left(\frac{r}{R}\right)^{3-p_0}\;.
\label{eq:mr}
\end{equation}
Note that a steeper volume density profile shows up as a shallower $r-m(r)$ relation (e.g. in Fig.~\ref{fig:rm3d}, the gas tracks are steeper than their star analogs).  The dash-dotted (black) straightline has a slope of 0.5, as expected for the stellar component in the low-density regime when $p_0=1.7$, i.e. $3-q \simeq 3 - 3p_0/2=0.5$.  While it indeed fits well the model with $p_0=1.7$ in the clump outer regions (open circles at $r \gtrsim 1$\,pc), the deviation increases towards the clump centre as the stellar density profile gets shallower than $q=3p_0/2$ (see middle panel of Fig.~\ref{fig:rhoprof}).  The lines with open symbols provide a direct mapping of how much stellar mass is enclosed within  a given radius.  For instance, when $p_0 = 1.7$ and $t=2$\,Myr, the embedded cluster has a radius $r_{ecl} \simeq 1$\,pc and a mass $m_{ecl} \simeq 400\,\Ms$, while the stellar mass enclosed within $R \simeq 6$\,pc is $\gtrsim 10^3\,\Ms$.    

\begin{figure}
\includegraphics[width=80mm]{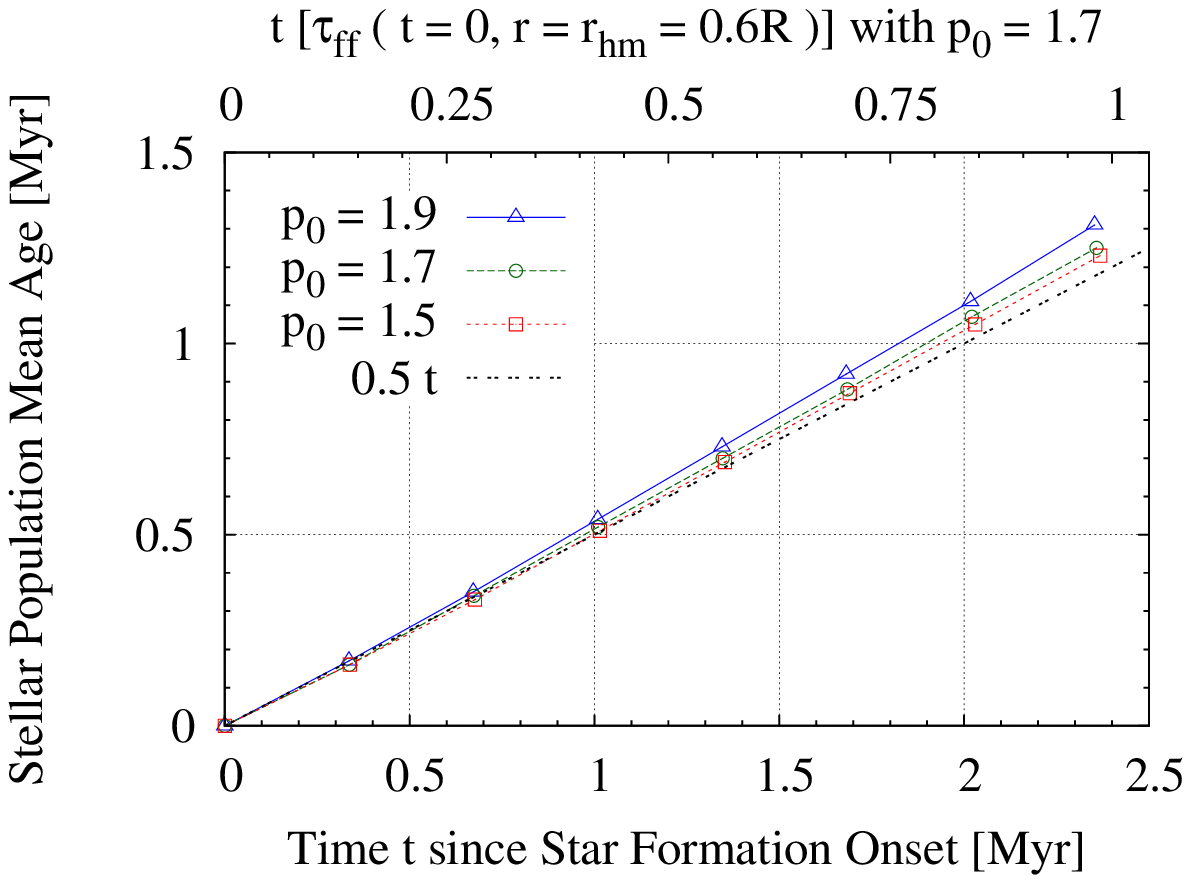}
\caption{Relation between the time $t$ elapsed since the onset of star formation in the simulations and the mean age of the stellar component.  Owing to the ongoing gas depletion and the star formation slow down it induces, the mean age is slightly older than half the simulation time-span.  The bottom $x$-axis shows the physical time (i.e. time in units of Myr) while the top $x$-axis shows time in units of the initial local free-fall time at the clump half-mass radius, $\tau_{ff}(t=0,r_{hm})$ when $p_0=1.7$.    \label{fig:tage} }
\end{figure}

Finally, we note that the time $t$ of our model corresponds to the {\it age-spread} of the model stellar component.  It differs from the mean age of the stars which would be inferred by an observer.  Figure~\ref{fig:tage} illustrates the mass-weighted {\it mean age} of the stars against the time-span $t$ of the simulations.  In case of a constant star formation rate, the mean age is simply $0.5t$.  However, in our model, star formation slows down with time and  the first stars to form (i.e. the oldest ones) tend to dominate the mass budget.  The mean age is therefore higher than $0.5t$, although the deviation is moderate, at most 10\% of half the simulation time-span.  The effect is stronger for $p_0=1.9$ than for $p_0=1.5$ since a steeper gas density gradient slows down star formation more than a shallow one (see Section \ref{ssubsec:glob} and Fig.~\ref{fig:sfet} below).

The top $x$-axis of Fig.~\ref{fig:tage} shows the time $t$ since star formation onset in units of the initial local free-fall time at the clump half-mass radius, $r_{hm}$, when $p_0=1.7$.  Using Eq.~\ref{eq:mr}, we find $r_{hm} \simeq 0.6R$ for $p_0=1.7$, which leads to an initial local density $\rho_0(r_{hm})\simeq 12 \Ms \cdot pc^{-3}$ (Eq.~\ref{eq:rho0}) and an initial local free-fall time $\tff(t=0,r_{hm}) \simeq 2.4$\,Myr.  A scaling in units of the initial free-fall time enables us to apply our model to molecular clumps with different mean densities and, therefore, different rates of star formation (since the mass of newly formed stars depends on the {\it ratio} $t/\tff(t=0,r)$; see e.g. Eq.~\ref{eq:mstap}).  For instance, a molecular clump that is 100 times denser initially forms stars at a rate that is 10-times faster (i.e. the bottom $x$-axis of Fig.~\ref{fig:tage} shrinks by a factor of 10).  This density-dependent star formation rate is akin to the density-dependent rate of dynamical evolution of star clusters after gas expulsion studied in depth by \citet{par12a}.

% --------------------------------
\section{Model Consequences}
\label{sec:conseq}
% --------------------------------

% ................................
\subsection{Star Formation Efficiencies: Global and Local}
\label{subsec:sfe}
%..................................

% ................................
\subsubsection{Global Star Formation Efficiency}
\label{ssubsec:glob}
%..................................

\begin{figure}
\includegraphics[width=80mm]{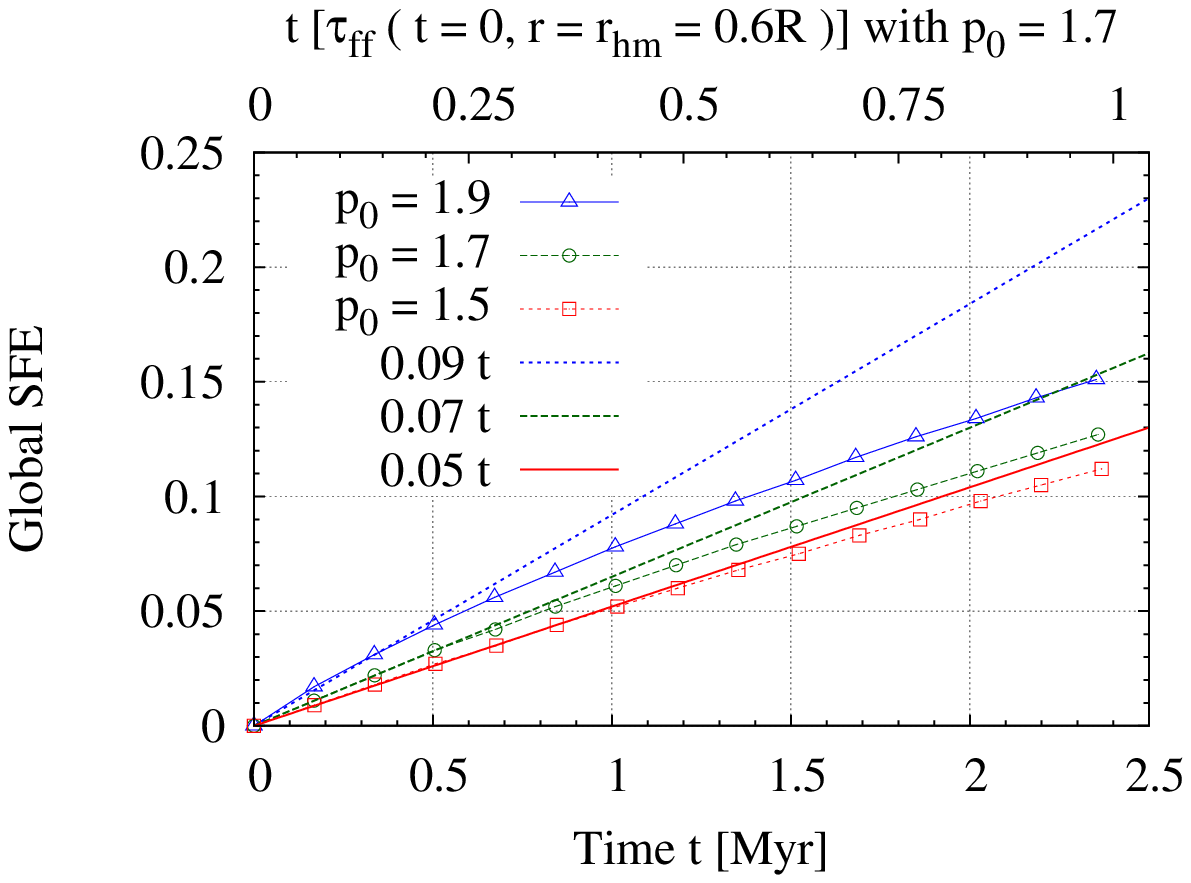}
\caption{Global star formation efficiency, SFE, in dependence of the time elapsed since the onset of star formation, $t$, for different initial density indices of the parent molecular clump, $p_0$ (see key).  A steeper profile (higher $p_0$) leads to larger star formation efficiencies at a given time $t$ because a higher gas mass fraction is located in the high-density central regions where the free-fall time is shorter.  The symbol-free lines illustrate the tangents to the numerical models at $t=0$. Units of $x$-axes as in Fig.~\ref{fig:tage}   \label{fig:sfet} }
\end{figure}

Figure \ref{fig:sfet} shows the evolution with time of the global star formation efficiency, SFE, namely, the stellar mass fraction averaged over the whole molecular clump: $SFE = M\st/M_0$.  The density indices of the precursor clump are identical to those in Fig.~\ref{fig:rhoprof}, i.e. $p_0=1.5, 1.7$ and $1.9$ (see key).  
The symbol-free lines depict the tangents to the models at the onset of star formation.  A steeper density profile concentrates a higher gas mass fraction into the clump inner regions, which have higher densities and shorter free-fall times.  This accelerates star formation initially (i.e. the tangent for $p_0=1.9$ is steeper than for $p_0=1.5$) and leads to higher global star formation efficiencies \citep[see also][]{tan06}.  With the ongoing gas depletion, star formation slows down as the comparison between the symbol-free lines and the numerical models illustrates.  
We note that, by a time $t=2.5$\,Myr, the global SFE remains low, of order 0.1-0.15.  We will discuss this result in the framework of star cluster survival in Section \ref{subsec:fb}.  We also remind that the predicted SFEs are parameter-dependent, here obtained with $M_0 = 10^4\,\Ms$, $R = 6\,$pc and $\eff = 0.1$.  Higher clump masses or smaller radii would lead to higher star formation efficiencies at a given time t.  

% ................................
\subsubsection{Radially-Varying Local Star Formation Efficiencies: Two- and Three-Dimensional}
\label{ssubsec:loc}
\begin{figure}
\includegraphics[width=80mm]{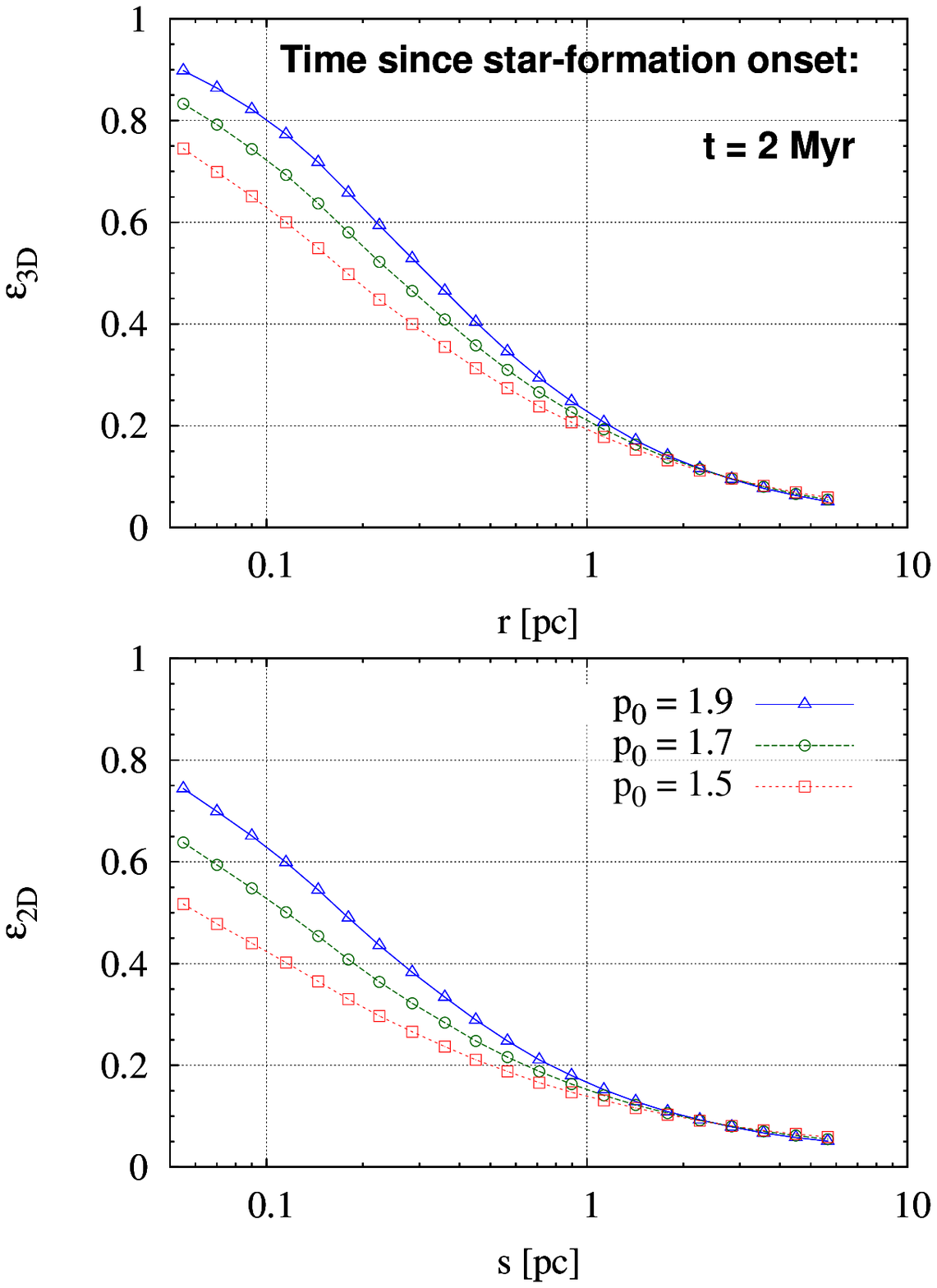}
\caption{{\it Top panel:} Three-dimensional local star formation efficiency, $\epsilon_{3D}$, as a function of radius $r$ (Eq.~\ref{eq:sfe3d}) for different density indices $p_0$ (see key in bottom panel). The progenitor clump of the star-forming region is the same as previously, i.e. $M_0 = 10^4\,\Ms$, $R = 6\,$pc and $\eff = 0.1$.  The time elapsed since the onset of star-formation is $t = 2$\,Myr.  {\it Bottom panel:}  Same for the two-dimensional star formation efficiency, $\epsilon_{2D}$, as a function of projected radius $s$ (Eq.~\ref{eq:sfe2d}) \label{fig:sfer} }
\end{figure}

%..................................
To define the three-dimensional local \sfe at the distance $r$ from the clump centre, we build on the local volume densities of stars and gas: 
\begin{equation}
\epsilon_{3D}(r) = \frac{\rhost(r)}{\rhost(r) + \rho_g(r)} = \frac{\rhost(r)}{\rho_0(r)}\;.
\label{eq:sfe3d}
\end{equation}
We remind the reader that the subscript `0' refers to the gas clump prior to star formation.
Equation \ref{eq:sfe3d} is illustrated in the top panel of Fig.~\ref{fig:sfer} for the same model as previously at a time  $t=2$\,Myr (symbol/colour-coding identical to Fig.~\ref{fig:sfet}).  As already introduced through the top panel of Fig.~\ref{fig:rhoprof}, the local \sfe is higher in the clump centre than in its outskirts.  It is worth keeping in mind that the derived efficiencies depend on the adopted \sfe per free-fall time, on the assumed duration of the star-formation process, and on the clump mean volume density.  That is, a higher $\epsilon_{ff}$, longer time $t$ and/or a larger clump mean volume density $3 M_0/(4 \pi R^3)$ would all increase $\epsilon_{3D}(r)$.  Also note that steeper density profiles achieve higher values of $\epsilon_{3D}$ since the clump central density is then higher.

$\epsilon_{3D}(r)$ is not the \sfe inferred by observers, however, since observers work with surface densities rather than volume densities.  We therefore define a two-dimensional \sfe based on the observed surface densities:
\begin{equation}
\epsilon_{2D}(s) = \frac{\Sigst(s)}{\Sigst(s) + \Sigma_g(s)} = \frac{\Sigst(s)}{\Sigma_0(s)}\;.
\label{eq:sfe2d}
\end{equation}
Equation \ref{eq:sfe2d} is shown in the bottom panel of Fig.~\ref{fig:sfer}.  

\begin{figure}
\includegraphics[width=80mm]{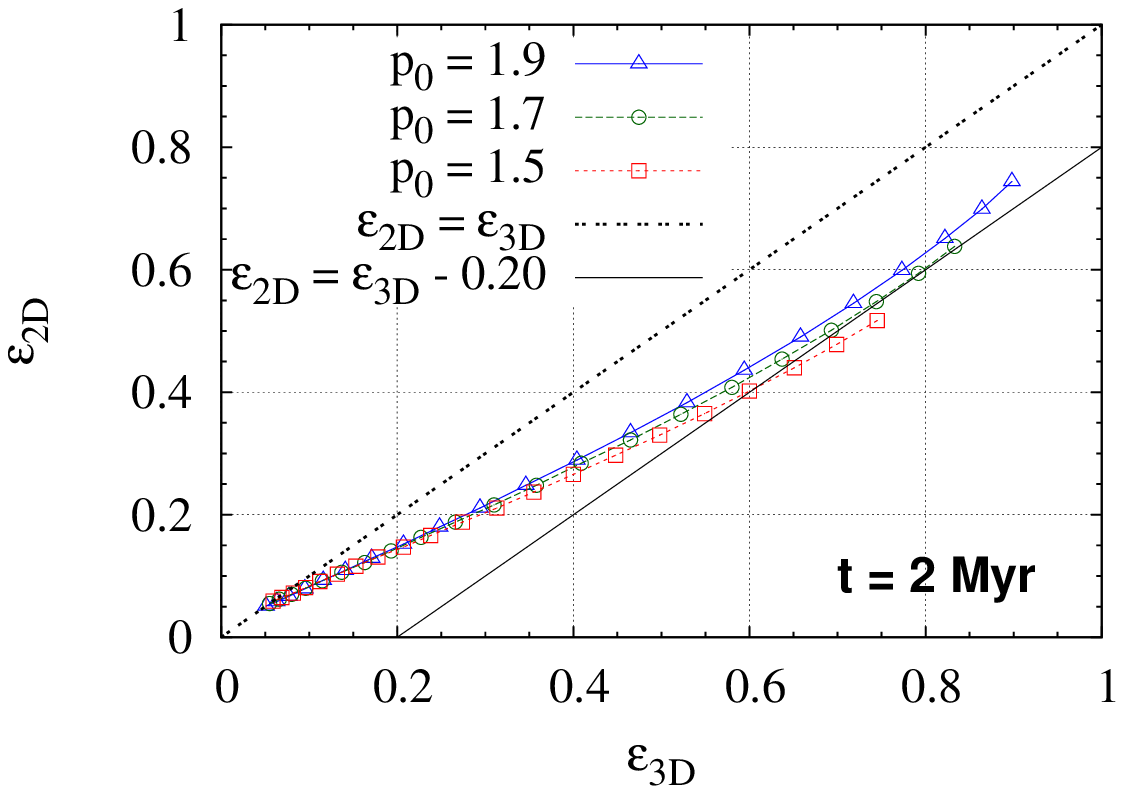}
\caption{Two-dimensional star formation efficiency, $\epsilon_{2D}$ (Eq.~\ref{eq:sfe2d}) against its three-dimensional counterpart, $\epsilon_{3D}$ (Eq.~\ref{eq:sfe3d}).  The (black) dotted and solid lines correspond to $\epsilon_{2D}=\epsilon_{3D}$ and $\epsilon_{2D}=\epsilon_{3D}-0.20$.  $\epsilon_{2D}$, the observed efficiency, is lower than $\epsilon_{3D}$, the actual efficiency, because surface densities measured around the clump centre intercept clump outskirts where the gas volume density and $\epsilon_{3D}$ are smaller     \label{fig:sfe2} }
\end{figure}

At a given radius $s=r$, $\epsilon_{2D}(s)$ is lower than $\epsilon_{3D}(r)$.  This is so because the star surface densities in the clump centre vicinity intercept outskirt material where the initial gas volume densities and achieved star formation efficiencies are lower.
This is further illustrated in Fig.~\ref{fig:sfe2} which depicts $\epsilon_{2D}$ in dependence of $\epsilon_{3D}$.  The dotted (black) line corresponds to $\epsilon_{2D}=\epsilon_{3D}$, while the solid one obeys $\epsilon_{2D}=\epsilon_{3D}-0.20$.  The solid line thus shows that, for the case of relevance here, the difference between the -- measured -- two-dimensional and -- actual -- three-dimensional star formation efficiencies can be as high as 20 per cent.

Both $\epsilon_{3D}$ and $\epsilon_{2D}$ rise from a few per cent at the clump edge up to higher than 50 per cent at the clump centre.  In particular, star formation depletes almost the entirety of the initial gas content in the clump centre and $\epsilon_{3D}$ reaches values as high as 60-90\,\% in the inner $r \lesssim 0.1$\,pc.  
This contrasts with the low {\it global} star formation efficiency, $SFE$, which we saw in Fig.~\ref{fig:sfet}.  This has important consequences for the survivability of (part of) the stellar component as a bound cluster after residual gas expulsion, as we shall discuss in Section \ref{subsec:fb}.  
The high central star formation efficiencies also allow us to propose an alternative explanation as to the low gas content in the central regions of the most evolved molecular clumps of \citet{hig09}.  According to \citet{hig09}, this absence or scarcity of gas results from its ongoing dispersal.  In contrast, in our picture, the absence of gas in the central regions of developed embedded clusters (e.g. the Orion Nebula Cluster) or of \citet{hig09} clumps stems from most of it having been fed to star formation, {\it even before gas dispersal starts}.

\begin{figure}
\includegraphics[width=80mm]{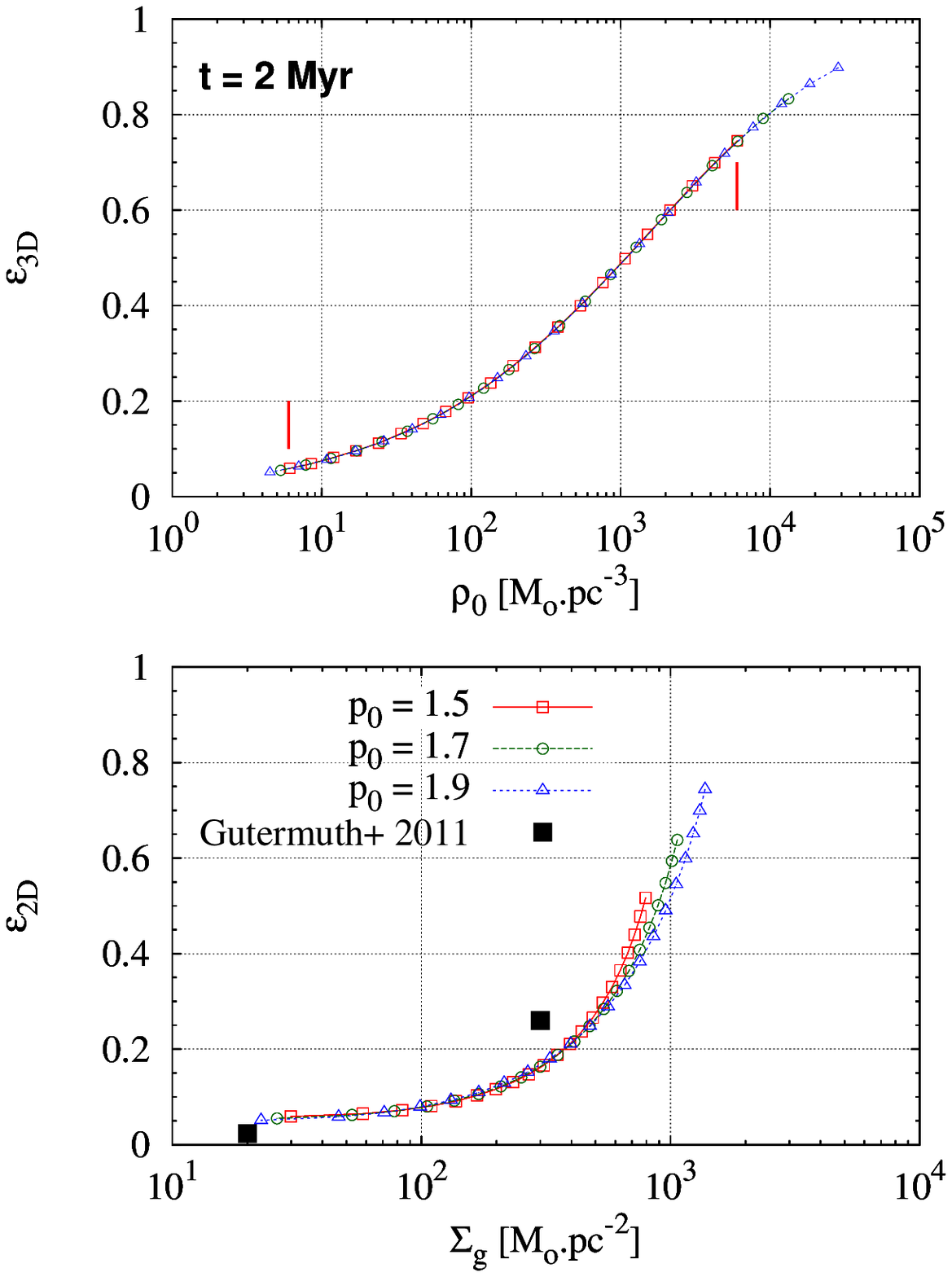}
\caption{{\it Top panel:} Three-dimensional star formation efficiency, $\epsilon_{3D}$, in dependence of the initial gas volume density, $\rho_0$.  The clump edge is on the left, the clump centre on the right.  The two vertical lines mark the limits of the $p_0=1.5$ model.  {\it Bottom panel:}  Two-dimensional star formation efficiency, $\epsilon_{2D}$, in dependence of the residual gas surface density, $\Sigma_g$.  The plain squares depict the measurements of \citet{gut11}.  In both panels, the models are identical to those in Figs.~\ref{fig:sfer} and \ref{fig:sfe2} \label{fig:sfedens} }
\end{figure}

The top panel of Fig.~\ref{fig:sfedens} shows how the three-dimensional \sfe increases from a few per cent to almost unity as the initial gas volume density rises from a few $\Ms \cdot pc^{-3}$ (near the clump edge) to higher than $10^4\,\Ms \cdot pc^{-3}$ (toward the clump centre).  We stress again that other model parameters will be conducive to different efficiencies.  With  $\rho_0 \simeq 7\,\Ms \cdot pc^{-3} \equiv n_{\rm H_2} \simeq 100\,cm^{-3}$ and $\rho_0 > 7000\,\Ms \cdot pc^{-3} \equiv n_{\rm H_2} > 10^5\,cm^{-3}$, where $n_{\rm H_2}$ is the molecular hydrogen number density, the range of volume densities characterizing the molecular clump under scrutiny here extends from the diffuse molecular gas probed in C$^{12}$O up to the dense gas traced in CS \citep[e.g.][]{shi03} or dust continuum emission \citep[e.g.][]{sch09}.   

The bottom panel of Fig.~\ref{fig:sfedens} illustrates the 2-dimensional \sfe in dependence of the residual gas surface density, that is, $\epsilon_{2D}$ versus $\Sigma_g$.  This is the relation which would be inferred from observing the star-forming region.  The two filled squares highlight the measurements made by \citet{gut11}: $\epsilon_{2D}=2.3$\,\% and $26$\,\% at $\Sigma_{g} = 20$ and $300\,\Ms\cdot pc^{-2}$, respectively.  They are practically consistent with our model despite it not being based on detailed star and gas density profiles.  It is interesting to note that star formation efficiencies of order 1\% happen at gas surface densities only slightly higher than the limit beyond which the gas becomes entirely molecular in spiral galaxies, i.e. $\Sigma=9\,\Ms\cdot pc^{-2}$ \citep[see top left panel in fig.~8 of][]{big08}.  Therefore, the lowest star formation efficiencies would correspond to the lowest possible surface densities where the gas is {\it entirely molecular}. [But see \citet{lad10}]  \\

\begin{figure}
\includegraphics[width=80mm]{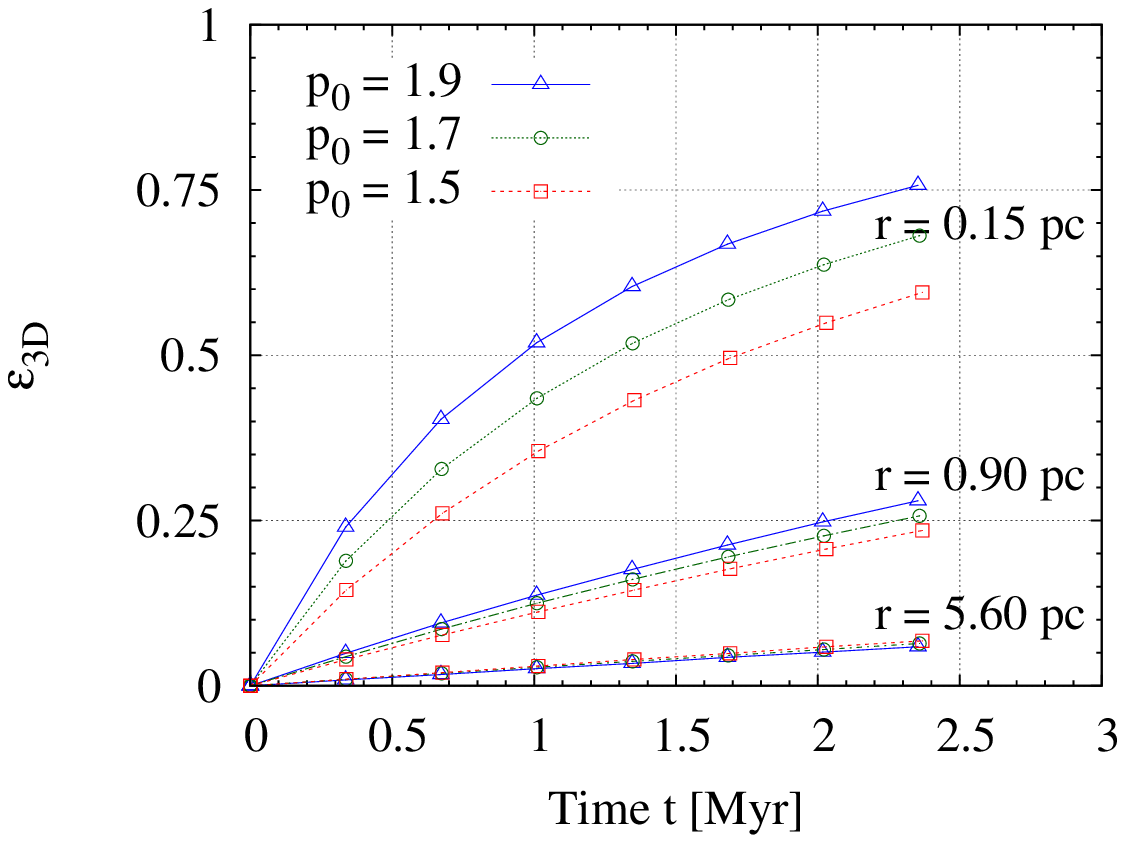}
\caption{Time evolution of the local star formation efficiency $\epsilon_{3D}$ for three different distances $r$ from the clump centre (see labels on the right) and three gas initial density indices $p_0$ (see key).  At the low density of the clump outskirts (e.g. $r=5.6$\,pc), the small gas depletion is conducive to $\epsilon_{3D}$ increasing linearly.  In contrast, at $r=0.15$\,pc, the strong gas depletion slows down star formation and the local star formation efficiency evolution eventually tends to saturate    \label{fig:lsfet} }
\end{figure}

Finally, Fig.~\ref{fig:lsfet} is the analog of Fig.~\ref{fig:sfet} for the local star formation efficiency, of which it shows the evolution with time for different radii $r$ and density indices $p_0$.  While the star formation rate is nearly constant in the clump outskirts (i.e. the local star formation efficiency is linearly increasing), the severe gas depletion in the clump inner regions slows down star formation there.  
 
% ....................................................................................
\subsection{Radially-Varying Local Star Formation Efficiency and Cluster Survival}
\label{subsec:fb}
%.....................................................................................
\begin{figure}
\includegraphics[width=80mm]{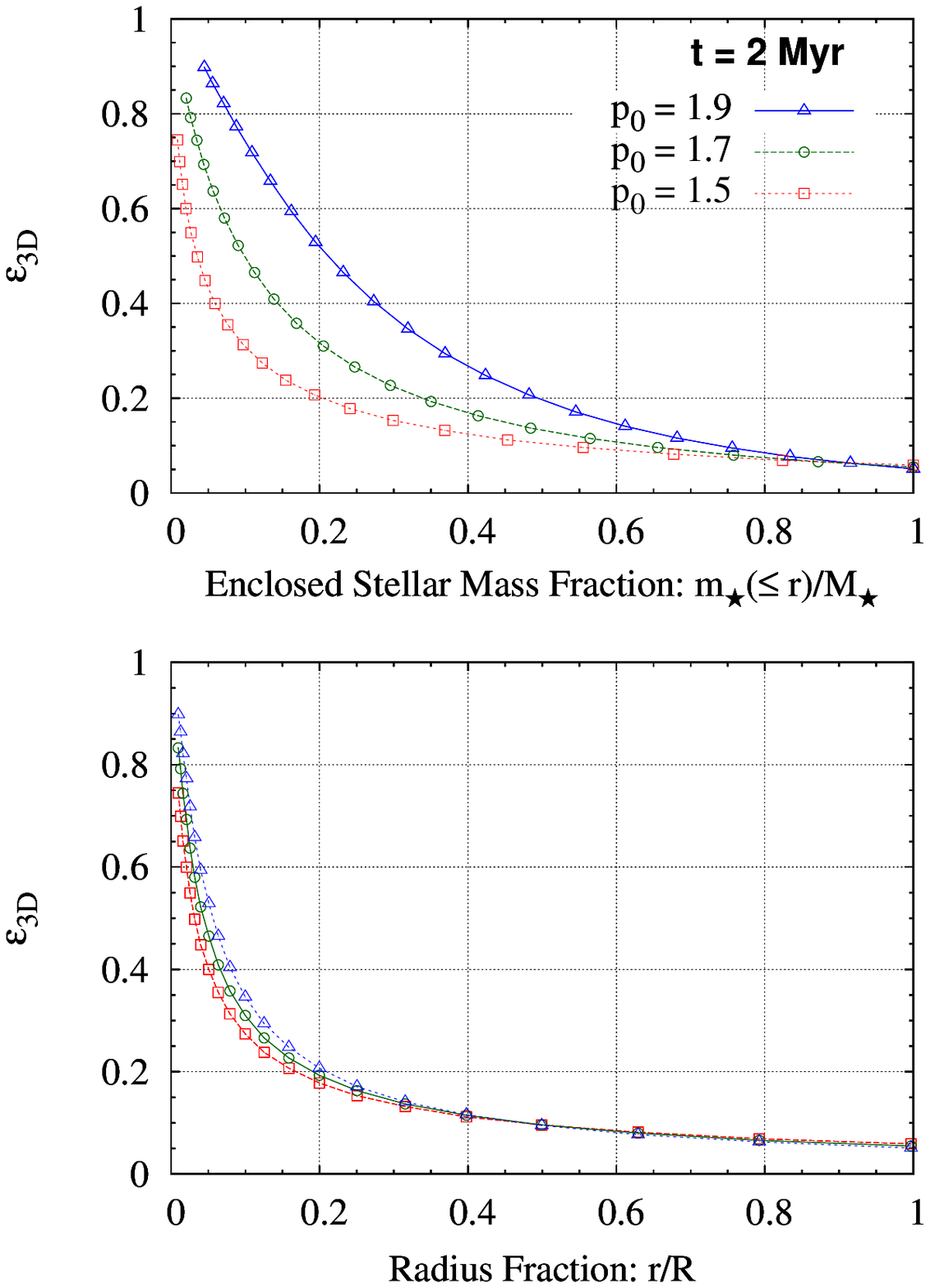}
\caption{{\it Top panel: } Relations between the local three-dimensional \sfe, $\epsilon_{3D}(r)$, and the enclosed stellar mass fraction, $m_{\star}(\leq r)/M_{\star}$.  {\it Bottom panel: } $\epsilon_{3D}(r)$ against normalized radius, $r/R$, where $R$ is the outer radius of the star-forming region. Symbol/colour-coding identical to Figs.~\ref{fig:sfet}-\ref{fig:lsfet}   \label{fig:enclm} }
\end{figure}

\begin{figure}
\includegraphics[width=80mm]{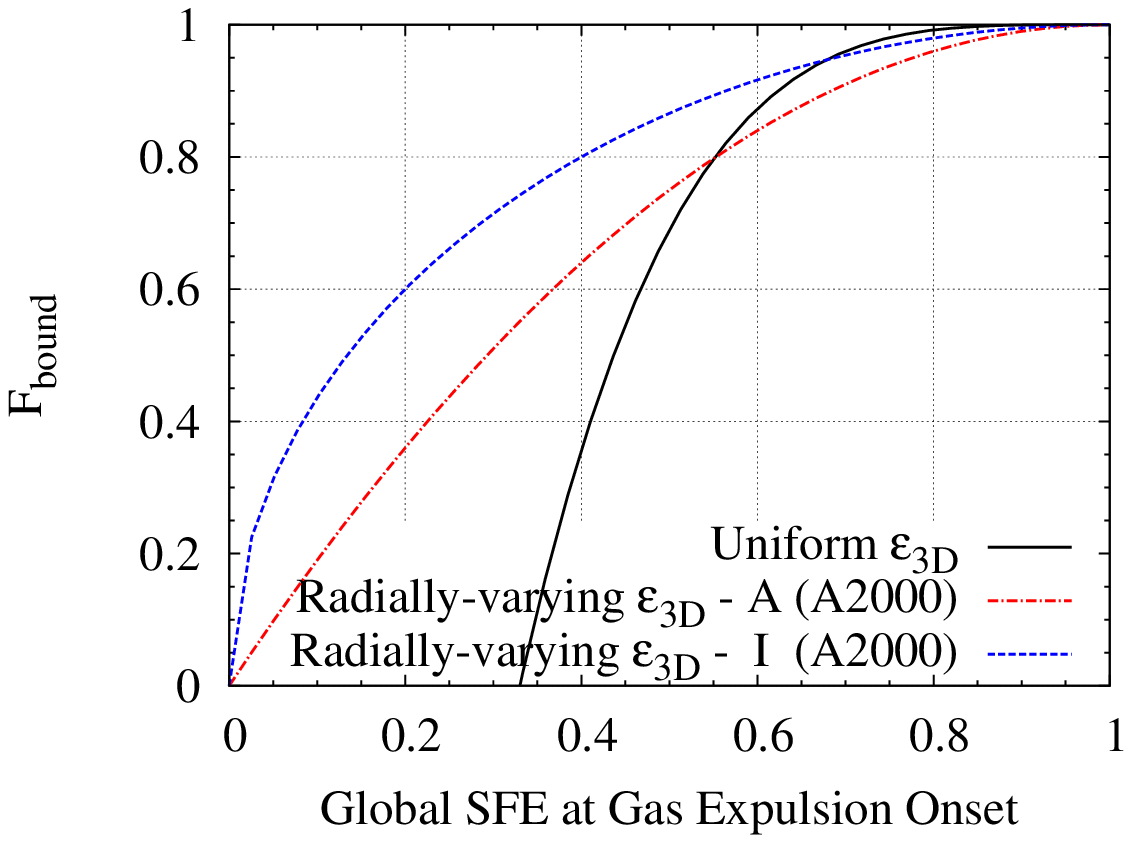}
\caption{Relation between the mass fraction of the stellar component forming a bound cluster after violent relaxation, $F_{bound}$, and the {\it global} SFE of the star-forming region.  The solid (black) line corresponds to a uniform local \sfe (i.e. $\epsilon_{3D}(r)=SFE$ irrespective of $r$).  The dashed and dash-dotted lines (blue and red) depict the models with radially-varying star formation efficiencies of \citet{ada00} (i.e. $\epsilon_{3D}(r)$ higher at smaller $r$ than at large $r$).  They build on either an isotropic (`I', see key) or anisotropic (`A') stellar velocity distribution.  In the low-SFE regime, the radially-varying models predict the survival of a bound cluster after violent relaxation ($F_{bound} > 0$) while the uniform model predicts the full disruption of the stellar component ($F_{bound}=0$).      \label{fig:fb} }
\end{figure}

Figure \ref{fig:sfet} shows that at the end of our simulations, the \sfe averaged over the whole star-forming region remains low, SFE $\simeq 10$-$15$\%.  At first glance, this is significantly smaller than the threshold required for the stellar component to retain a bound star cluster after residual gas expulsion.  The relation between SFE and the fraction of stars remaining bound to the stellar component, $F_{bound}$, is shown in fig.~1 of \citet{par07} based on the simulations of \citet{gey01} and \citet{bau07}.  For the sake of clarity, this $F_{bound}$-vs-SFE relation is reproduced as the solid (black) line in our  Fig.~\ref{fig:fb}.  It assumes instantaneous gas expulsion and a weak external tidal field.  The threshold to retain a bound group of stars ($F_{bound} > 0$) is $SFE_{th} \simeq 0.33$.  We stress that this threshold necessarily depends on the hypotheses of the models used to derive it.  The vast majority of cluster-gas-expulsion models developed so far assume that the local \sfe is uniform all through the star-forming region, that is, $\epsilon_{3D}$ is independent of $r$ and constant (as a result SFE=$\epsilon_{3D}$).  This is in stark contrast with our result of Section \ref{subsec:sfe} that the local \sfe spans near to two orders of magnitude, from almost unity down to a few per cent (top panel of Fig.~\ref{fig:sfer}).  This property of the star-forming region has important implications for the ability of the stellar component to retain a bound star cluster after gas expulsion.
Owing to its small residual gas fraction, the clump central region has a high resilience to gas expulsion. It will therefore necessarily produce a bound cluster.  Most of the gas-expulsion-driven disruption of the stellar component takes place in the outskirts where the local \sfe is low.  

The top panel of Fig.~\ref{fig:enclm} shows the local \sfe in dependence of the enclosed stellar mass fraction.  The bottom panel depicts $\epsilon_{3D}$ against the normalized radius $r/R$ (i.e. another representation of the top panel of Fig.~\ref{fig:sfer}, with the $x$-axis scale now linear).  
Let us consider the model with a density index $p_0 = 1.9$ (blue line with open triangles).  About 15\% of the stellar mass arises from gas having experienced a local \sfe $\epsilon_{3D} \geq 0.60$.  Under the assumption of a uniform \sfe in that limited central region ($0.60 \leq \epsilon_{3D} \leq 0.95$ when $r/R \leq 0.04$ or $r \leq 0.25$\,pc; see bottom panel of Fig.~\ref{fig:enclm}), we can use the solid black line in Fig.~\ref{fig:fb} to infer the bound fraction of stars after violent relaxation.  
More than 85\% of the stellar mass remains bound in the central region.  We can now derive a lower limit on the mass of a surviving bound cluster.  With a global SFE of $\simeq$ 13\% at a time $t=2$\,Myr (Fig.~\ref{fig:sfet}), the whole stellar component has a mass $M\st \simeq 1300\,\Ms$, of which 15\% are located in the central area practically unaffected by gas expulsion.  Therefore, a cluster of at least 200\,$\Ms$ will remain after violent relaxation in spite of the low global SFE.  In other words, over the global scale of the whole star-forming region, the bound fraction $F_{bound}$ must be at least $15$\%.  
 
\citet{ada00} developed a pioneering model accounting for a radially-dependent local star formation efficiency, which quantifies the effect introduced above.  That is, because of the high {\it local} \sfe in the central regions (i.e. high $\epsilon_{3D}$ at small $r$), a cluster survives gas expulsion ($F_{bound} > 0$) despite a low {\it global} \sfe (low SFE).  The relations he derived between SFE and $F_{bound}$ are shown in Fig.~\ref{fig:fb} as the dash-dotted (red) and dashed (blue) lines.  They  correspond to isotropic and anisotropic stellar velocity distributions, respectively.  For SFE $\simeq 0.1$-$0.15$, uniform-$\epsilon_{3D}$ models predict the stellar component full disruption (solid black line: $F_{bound}=0$).  In contrast,  the radially-varying models of \citet{ada00} give bound fractions as high as 20-50\%.  With $M_0 = 10^4\,\Ms$, SFE$=0.13$ and $F_{bound} \simeq 0.3$, our star-forming region would thus give rise to a bound star cluster with a mass: $M_{cl} = F_{bound} \cdot SFE \cdot M_0 \simeq 400\,\Ms$.  This is in fair agreement with the lower limit of $200\,\Ms$ we derived above.

% ---------------------------------------------------------------
\section{Discussion and Future Work}
\label{sec:fut}
% ---------------------------------------------------------------

In this section, we firstly discuss how our results and those from hydrodynamics simulations of star formation relate to each other.  We then touch on two aspects of our work which we will refine in the near future, namely, the value of the \sfe per free-fall time ($\eff$), and the impact of the density profile of molecular clumps.

%............................................
\subsection{Clump density profile and embedded-cluster morphology}
\label{ssec:lit}
%............................................

\citet{mue02} detected in their sample of star-forming molecular clumps a correlation between the clump radial density profile and mass, with higher-mass clumps being on the average steeper (top panels of their fig.~18).  The observed $p$-index range ($0.7 \lesssim p \lesssim 2.5$) can now be related to the comprehensive set of adaptative-mesh refinement simulations performed by \citet{gir11}.  These simulations show that the initial gas density gradient of spherical molecular clumps is of paramount importance to the morphology of the star clusters they form.  
In clumps initially uniform in density, supersonic turbulence has the time to compress the gas in locally disconnected areas before global collapse sets in.  This results in locally disconnected filaments and spatially distinct subclusters of sink particles.  In contrast, centrally concentrated clumps (e.g. a Bonnor-Ebert or power-law $p_0=1.5$ density profile) form filaments more strongly connected and centrally concentrated because of the shorter time-scale for global collapse in the clump central region.  This eventually favours the formation of one main central cluster \citep[see fig.~4 in ][]{gir11}.  The morphology of these `single-block' clusters -- substructured or smooth power-law density profile -- depends on the initial turbulent velocity field \citep{gir12b}.  Solenoidal turbulence in Bonner-Ebert or $p_0=1.5$ density profiles can produce embedded clusters with a smooth radial density profile.  For instance, the embedded-cluster $Q$-parameter
\footnote{The $Q$-parameter \citep[][]{cart04} quantifies the degree of substructures in star clusters.  It is defined as the ratio between the normalised mean separation of the stars and the normalised mean length of the edges of the minimal spanning tree associated to the (proto)cluster.  A cluster with $Q<0.8$ is substructured/fractal, while $Q>0.8$ indicates a smooth radial density profile} 
can be of order 1.2 \citep[figs~11 and 12 in][]{gir12b}.  We note that $0.9 \lesssim Q \lesssim 1.5$ corresponds to a power-law density index $2 \lesssim q \lesssim 2.9$ for the embedded cluster \citep[see Table 1 in][]{cart04}, a $q$-range which agrees well with our model predictions (see our Eq.~\ref{eq:uplim}: $q \lesssim 3p_0/2=2.25$ if $p_0=1.5$).  We caution, however, that our model does not account for the clump global collapse and are therefore not fully comparable to those of \citet{gir11}.  Finally, we note that these adaptative-mesh refinement  simulations predict that the first stars form in the clump central region \citep{gir12a}.  Once secondary stars form around the central objects, they accrete the infalling gas preventing it from reaching the clump central region (`fragmentation-induced starvation').  This forces new protostars to form at increasingly large distance from the clump centre, an aspect which agrees well with \citet{pfa11} 's scenario. \\

If the correlation between the clump density index and mass found by \citet{mue02} is genuine (a bias arising from more massive sources being located at larger distances where the resolution is poorer cannot be fully excluded yet), one would expect shallow low-mass clumps to produce ensembles of subclusters, while steeper high-mass clumps would give rise to `single-block' clusters.
Which mode of star cluster formation -- subclustered or `single-block' -- dominates the process of star formation then depends on the slope of the clump mass function.  Observed mass function slopes are shallower than $-2$ \citep[e.g. $\simeq -1.7$, ][]{kra98}.  High-mass clumps therefore dominate the star-forming gas mass and most newly formed stars may thus originate from `single-block' clusters.  Low-mass clumps and subcluster ensembles  would dominate the cluster population only in terms of number.  

Similarly to the uniform density profile of \citet{gir11}, the end product of the hydrodynamics simulations of  \citet{bon08} is an ensemble of subclusters \citep[see ][for their merging history]{mas10}.  This is not surprising since the gas cylinder they scrutinize is characterized by a weak density gradient along its main axis.  Despite a low global SFE (SFE $\simeq 0.15$), their subclusters are gas-poor on a length scale of 0.1-0.2\,pc, an effect due to the accretion of gas onto sink particles and the accretion-induced subcluster shrinkage \citep{kru12}.  This is reminiscent of the locally high SFE achieved by the central region of our molecular clump despite a globally low one (Figs \ref{fig:sfet} and \ref{fig:sfer}).  Note, however, that the respective embedded-cluster morphologies are different (subclustered in \citet{bon08} vs. a smooth radial density profile in our model).  Given their high local SFE and  small number of stars, the relaxation time of subclusters is short enough for their early dynamical evolution to be dominated by  collisional stellar dynamics rather than violent relaxation.  As a result, \citet{moe12} find that  subclusters expand due to the scattering of their stars, eventually erasing the subcluster structure \citep[see also][]{smi08}.  We remind that these results hold for the low-mass end of the cluster mass spectrum (i.e. cluster mass of the order of $100\,\Ms$), typical of the Solar Neighbourhood.

Finally, we emphasize that the outcomes of smooth-particle hydrodynamics and adaptative mesh refinement simulations of star-forming regions necessarily depend on their input physics.  It remains ill-known how the inclusion of magnetic fields or radiative feedback (hence gas heating), how treating the protostars as extended gas spheres rather than sink particles, would
affect the emerging properties of model star-forming regions.  Additionally, these simulations do not currently access the regime of ab initio formation of high-mass clusters.  This is up to at least two orders of magnitude higher than the gas  and stellar masses covered so far.  How relevant to the formation of high-mass clusters the physical processes at work inside low-mass clouds are remains an open question.

%............................................
\subsection{The \sfe per free-fall time}
\label{ssec:eff}
%............................................

To match the observed local star formation law of \citet{gut11} with our model ($M_0 \simeq 10^4\,\Ms$, $R \simeq 6$\,pc) at $t=2$\,Myr requires a \sfe per free-fall time of $\eff = 0.1$.  We insist that the so-derived value of $\eff$ is parameter-dependent.  A longer time $t$ or a higher clump density (i.e. shorter free-fall time) would both lead to higher YSO surface and volume  densities.  That would require a smaller $\eff$ to maintain the good match between the model and the scaling law averaging the observations shown in the top panel of Fig.~\ref{fig:sfl}.

At this stage, it is therefore premature to draw any conclusion about our $\eff$ estimate being an order of magnitude higher than that derived by \citet{kru07}.  \citet{kru07} consider the fraction of the Milky Way star formation rate taking place in different classes of objects, from Giant Molecular Clouds (number densities $n_{\rm H_2} \simeq 100\,cm^{-3}$) to HCN-traced molecular clumps ($n_{\rm H_2} \simeq 6 \cdot 10^4\,cm^{-3}$).  Building on the total mass of these objects in the Galaxy and on their mean density, they derive a \sfe per free-fall time of about one per cent, independent of volume density (SFR$_{ff} \simeq 0.01$ in their notation; see their eq.~1 and fig.~5).  
On the other hand, in a study of {\it individual} molecular clouds, \citet{eva09} infer $0.03 \leq {\rm SFR}_{ff} \leq 0.06$ (their section 4.3).  This is intermediate between the estimate of \citet{kru07} and ours.  

Ideally, to derive $\eff$ in the framework of our model, we need the total masses of gas and stars enclosed within a given radius (not simply orders of magnitude as used in this introductory paper), as well as the time elapsed since the onset of star formation.  In a forthcoming paper, we will apply our model to smaller-scale molecular clumps with known size, gas mass and star mass.  This will allow us to put on a firm footing the comparison between our $\eff$ estimate, those of \citet{eva09} and \citet{kru07}.  The comparison will be especially interesting as these three works define a sequence of distinct spatial scales, from Galaxy-integrated star formation \citep{kru07}, to star formation in individual molecular clouds \citep{eva09} and star formation in individual molecular clumps (our work).  

%............................................
\subsection{The time-evolving gas density profile of molecular clumps}
\label{ssec:prof}
%............................................

Summing up the observed surface density profiles of the stars and (unprocesssed) gas in star-forming regions will provide a direct estimate of $p_0$, the initial density index.  In that respect, our assumption of a single $p_0$-value valid over 6\,pc in radius is probably an oversimplification.  \citet{pir09} finds that the density index of massive-star-forming clumps is $p = 1.6 \pm 0.3$ within 0.8\,pc from their centre.  Beyond that distance, the density profile drops steeper.  \citet{beu02} reach a similar result for the massive star-forming regions they study.  Two power-laws are needed to describe their clump density profiles: $p = 1.6 \pm 0.5$ within 32\arcsec from the clump centre, and  a steeper density profile beyond (their fig.~2).  The break at 32\arcsec corresponds to half of their fittable range and equates with 1\,pc at the typical 6\,kpc distance of their star-forming regions.  Therefore, it will be interesting in the future to convert our present one-zone model into a two-zone one and to quantify how this affects the estimate of $\eff$.

In addition, it is worth noting that the density indices reported by \citet{beu02}, \citet{mue02} and \citet{pir09} refer to {\it star-forming} molecular clumps.  Yet, the bottom panel of Fig.~\ref{fig:rhoprof} shows that, as the stellar content builds up, the gas density profile gets shallower, again reflecting the faster gas-to-star conversion in the clump inner regions.  Over the radial range 0.2-1.0\,pc and over a time-span of 2.5\,Myr, this gas density profile evolves from $p_0 = 1.7$ to $p \simeq 1.3$.  It is therefore likely that observed star-forming molecular clumps had at star formation onset  steeper profiles than is observed now.  Models starting with $p_0 =1.9$ may thus be more appropriate than those with $p_0 =1.5$ or $p_0 =1.7$.  One way of testing this observationally would be to compare the density profiles of starless molecular clumps \citep[e.g.][]{tac12} and star-forming ones.  

% ---------------------------------------------------------------
\section{Summary and Conclusions}
\label{sec:conc}
% ---------------------------------------------------------------
We have presented a model quantifying the stellar content of molecular clumps as a function of time and initial gas volume density.  The model key-ingredient is the \sfe per free-fall time, $\eff$, namely, the mass fraction of gas turned into stars per free-fall time, $\tff$ (Eq.~\ref{eq:tff}).  The model originality resides in it building on a {\it local} free-fall time defined in relation to the {\it local} volume density of the star-forming gas.  That is, the radial volume density gradient of molecular clumps leads to a radially-dependent free-fall time, shorter in the clump central regions and longer in the clump outskirts.  In other words, star formation proceeds more quickly at shorter distance from the clump centre.  As a result of this differential rate of star formation, the radial density profile of the stellar population built by the clump (which we refer to as the `stellar component') is steeper than that of the clump initially (i.e. $q \lesssim 3p_0/2$, where $q$ and $p_0$ are, respectively, the density indices of the stellar component and of the clump initially; see Eq.~\ref{eq:uplim}).

We combine the volume density profile of spherical molecular clumps (Eq.~\ref{eq:rho0}) with their radially-dependent free-fall time (Eq.~\ref{eq:tff}) to model the density profile of the stellar component as a function of time and gas initial density.  This is done both numerically (Section \ref{subsec:num}) and analytically  (Section \ref{subsec:anly}).

We then project the volume density profiles for the unprocessed gas and stars and obtain the corresponding surface density profiles (Section \ref{subsec:sfl}).  Under the assumptions that YSOs do not migrate away from their birth sites significantly, and that molecular clumps do not experience significant gas motions and/or inflows and/or outflows, predicted surface density profiles are  compared to the observations.  Specifically, we consider the local star formation law inferred by \citet{gut11}.  It relates the local surface density of molecular gas, $\Sigma_{g}$, and the local surface density of YSOs, $\Sigma\st$: $\Sigma\st \simeq 10^{-3} \Sigma_{g}^{\alpha}$, with $\alpha \simeq 2$ and the surface densities in units of $\Ms \cdot pc^{-2}$.  We find that a stellar density profile steeper than a gas density profile, as predicted by our model (Eq.~\ref{eq:uplim0} and top panel of Fig.~\ref{fig:rhoprof}), naturally leads to $\alpha > 1$, as observed by \citet{gut11}.  In  particular, molecular clump density indices $p_0$ in the range $1.5 \leq p_0 \leq 1.9$ \citep{beu02, mue02} are conducive to $\alpha \simeq 2$, in excellent agreement with what is observed (top panel of Fig.~\ref{fig:sfl}).  

To compare the normalizations of the observed and predicted star formation laws, we consider the specific case of a molecular clump with an initial gas mass $M_0 \simeq 10^4\,\Ms$ and a radius $R \simeq 6$\,pc.  These mass and radius are chosen so as to emulate the most prominent concentration of YSOs in the MonR2 molecular cloud (Section \ref{subsec:num}).  We find that the predicted  star formation law agrees with its observed counterpart about $2$\,Myr after the onset of star formation (i.e. $t = 2$\,Myr) for a \sfe per free-fall time $\eff = 0.1$. 
For that same model, we derive the YSO surface-density distribution (Section \ref{subsec:Sigdist}; Fig.~\ref{fig:Sigdist}) and find a good agreement with the observed distribution of the Solar Neighbourhood inferred by \citet{bre10}.  We stress that this comparison is only preliminary given that our present model covers one single molecular clump, while the distribution of \citet{bre10} encompasses several star-forming regions.

In the second part of the paper, we combine our model of a time-evolving star-forming region with a surface density threshold (Section \ref{sec:ecl}).  Prior to the one-to-one identification of YSOs based on their infrared excess with the {\it Spitzer} telescope, the subtraction of a given surface density accounting for the contaminating foreground and background stars was often relied on to define the limiting radius of a cluster and its mass \citep[see section 2 of][for a summary of the different techniques for cluster member identification]{all07}.  The combination of our model predictions with a surface density threshold can thus be compared with pre-{\it Spitzer} cluster data sets, such as the star-cluster catalog compiled by \citet{lad03}.

\citet{pfa11} finds that the clusters of \citet{lad03} define a sequence of decreasing volume density along with increasing radius.  She suggests this sequence to be a time sequence corresponding to the growth of the cluster stellar content, an hypothesis we confirm.  As time goes by, the total stellar mass increase raises the fraction of the stellar component seen above the surface density threshold (we refer to this `emerging' part as the `embedded cluster').  Therefore, the observed  embedded-cluster radius gets larger too (red diamonds in the bottom panel of Fig.~\ref{fig:vh}).  The mean surface density of the embedded cluster remains approximately constant since the cluster definition rests on a surface density limit.  
A constant surface density equating with $\rho_{ecl} \propto r_{ecl}^{-1}$, the observed growth of the embedded-cluster radius thus leads to a decrease of the observed mean volume density.  In contrast, the outer radius of the stellar component as a whole stays constant and its mean volume density increases with time (see the sequences of red and blue circles in the top panel of Fig.~\ref{fig:eclseq}).

In Section \ref{sec:conseq}, we have discussed the impact of a radially-varying local \sfe on the post-gas-expulsion evolution of clusters.  That the density profile of the stellar component built by the clump is steeper than that of the gas initially implies a local \sfe increasing towards the clump centre.  As a result, the clump central region is more resilient to gas expulsion than the clump outskirts.  Despite a low {\it global} \sfe (SFE $\simeq$ 10--15\% at $t=2$\,Myr for $M_0 \simeq 10^4\,\Ms$, $R \simeq 6\,pc$ and $\eff = 0.1$, see Fig.~\ref{fig:sfet}), the stellar component can still leave a bound cluster after violent relaxation due to the high {\it local} star formation efficiency ($\epsilon_{3D}$) in the clump innerst regions (Fig.~\ref{fig:sfer}).  \citet{ada00} computed the bound fraction of stars at the end of violent relaxation for a  centrally-peaked \sfe profile.  He found that such a configuration increases the cluster survivability compared to models where the star formation efficiency is uniform all through the molecular clump (Fig.~\ref{fig:fb}).  

Although our model addresses the formation of leaky (i.e. low-density) clusters, it already provides a hint as to why some starburst (i.e. high-density) clusters may be in virial equilibrium despite their very young age \citep[e.g. Westerlund-1,][]{men07}.  The top panel of Fig.~\ref{fig:sfedens} shows that volume densities characteristic of starburst clusters (i.e. several $10^3\,\Ms \cdot pc^{-3}$ at least) lead to three-dimensional star formation efficiencies in excess of 50\% (for the star formation efficiency per free-fall time adopted here).  
A high \sfe limits the impact of gas expulsion on the cluster dynamics and the departure from virial equilibrium after gas expulsion.  [Besides, a high density also implies a short crossing-time which hastens the return of the cluster to equilibrium \citep{par12a}].  That does {\it not} imply that gas expulsion is not a major driver of young cluster dynamics in general.  Only high-density regions -- e.g. dense molecular clumps close to the Galactic centre and central regions of massive molecular clumps in the Galactic disc -- may be relatively immune to gas expulsion.  

Finally, we note that a sound quantification of the stellar component also requests an accurate estimate of the \sfe per free-fall time ($\eff$) since this parameter drives the relation between the local \sfe $\epsilon_{3D}$ and the local volume density $\rho_0(r)$ at a given time (top panel of Fig.~\ref{fig:sfedens}).  Should $\eff$ be halved, the predicted global and local star formation efficiencies will be reduced by a factor of two (at most).  Here arises a direct connection between the rate of star formation in molecular clumps and the stellar dynamics of post-gas-expulsion clusters.  That is, the rate of dissolution of young clusters and the \sfe per free-fall time are tightly related.

\section{Acknowledgments}
Throughout this work, G.P. has been supported by a Research Fellowship of the Max-Planck-Institut f\"ur Radioastronomie (Bonn), and an Olympia-Morata Fellowship of Heidelberg University.  
We thank Karl Menten for a careful reading of our manuscript, and the referee, Cathie Clarke, for a very constructive report.

\end{document}